\documentclass[12pt]{iopart}

\usepackage{iopams}  

\usepackage{amsgen}
\usepackage{amsfonts}
\usepackage{amsbsy}
\usepackage{amssymb}

\usepackage{setstack}

\usepackage{graphicx} 
\usepackage[biblabel]{cite}

\usepackage{xcolor}

\begin{document}

\title{Bosonic two-stroke heat engines with polynomial nonlinear coupling}

\author{G Chesi$^{1,2,*}$, C Macchiavello$^{1,2}$ and M F Sacchi$^{1,3}$}
\address{$^1$ QUIT Group, Dipartimento di Fisica, Universit\`a degli Studi di Pavia, Via Agostino Bassi 6, 27100 Pavia, Italy}
\address{$^2$ National Institute for Nuclear Physics, Sezione di Pavia, Via Agostino Bassi 6, 27100 Pavia, Italy}
\address{$^3$ CNR - Istituto di Fotonica e Nanotecnologie, Piazza Leonardo da Vinci 32, I-20133, Milano, Italy}
\address{$^*$ Author to whom any correspondence should be addressed.}
\ead{giovanni.chesi@unipv.it}



\begin{abstract}
We study the thermodynamics of two-stroke heat engines where two
bosonic modes $a$ and $b$ are coupled by the general nonlinear
interaction $V_{\theta} = \exp {(\theta a^{\dagger n}b^m - \theta^*
  a^nb^{\dagger m})}$. By adopting the two-point measurement scheme we
retrieve the distribution of the stochastic work, and hence the
relative fluctuations of the extracted work up to the second order in
the coupling $\theta$. We identify the optimal interactions providing
large average work with small fluctuations in the operational regime
of the heat engine. Then, we consider the specific cases $n=2$, $m=1$
and $n=1$, $m=2$ up to the fourth order in $\theta$. We optimize the
average work and the signal-to-noise ratio over the frequencies of the
bosonic modes and the temperatures of the reservoirs. Finally, we
determine the thermodynamic uncertainty relations for these processes
in relation with the order of the expansion of the unitary interaction
$V_{\theta}$.
\end{abstract}

%
%
%
%
%

\section{Introduction}
The main goal of a heat engine is the extraction of work from physical
systems by using thermal reservoirs at different temperatures. A quantum description is needed for engines operating with microscopic working systems, and the thermodynamics in this regime has been under investigation for the last decades \cite{ramsey1956,scovil1959,geusic1967,alicki1979,kosloff1984,deroeck2004,allahverdyan2004,talkner2007,talkner2007cf,quan2007,allahverdyan2008,esposito2009,campisi2011,bar,timpanaro2019,macchiavello2020,max,max2,ding2021,ergo,biswas2024}. In this context, the very definition of work is matter of debate \cite{campisi2011}. An interesting and useful definition has been given in terms of the maximum work extractable from finite systems and compatible with quantum mechanics, namely the \textit{ergotropy} \cite{allahverdyan2004}. In Ref.~\cite{allahverdyan2008}, the authors optimize a quantum heat engine for the extraction of the ergotropy. In particular, they consider two systems with finite dimension at equilibrium, each one coupled with the pertaining reservoir, at two different temperatures. Then, the two systems are disconnected from the reservoirs and coupled with each other through a unitary interaction extracting work. Finally, the interaction is turned off and the systems are re-connected to the respective reservoirs to reach equilibrium, thus closing the cycle. This scheme defines the so-called \textit{two-stroke Otto heat engine} \cite{quan2007}.
\\
However, the amount of extracted work is not the sole figure of merit, particularly in practical scenarios. Other key performance indicators include the signal-to-noise ratio (SNR) - or equivalently, the relative fluctuations (RF) of the work output \cite{timpanaro2019,max,max2,ergo} - and the operational range of the engine \cite{allahverdyan2004,ergo,biswas2024}. In two-stroke heat engines employing bosonic modes \cite{macchiavello2020,max} or qudits \cite{timpanaro2019,max2} as working media, the SNR of the work extracted by the swap interaction has been analyzed and linked to the mean entropy production through thermodynamic uncertainty relations (TUR) \cite{bar,timpanaro2019,max,max2,gingrich2016,pietzonka2017,proesmans2017,potts2019,hasegawa2019,vo2020,vanvu2022,salazar2022,francica2022,vanvu2023}. 
\\
The TURs considered in this work originate from the foundational inequality established in Ref.~\cite{bar} for steady-state systems. Subsequent studies have extended their applicability to finite-time processes \cite{pietzonka2017} and to some cyclic thermal machines \cite{timpanaro2019, max,max2,ergo}, and re-formulated  them in the context of Markov chains and periodically driven systems \cite{proesmans2017}. Notably, Ref.~\cite{max} derived a tighter TUR tailored to two-mode bosonic heat engines operating under a swap interaction, which constrains the fluctuations $\rm{var}(W)$ of the mean extracted work $\langle W \rangle$ as follows:
\begin{equation} \label{maxtur}
\frac{{\rm var}(W)}{\langle W\rangle^2} \geq \frac{2}{\langle \Sigma \rangle}+1,
\end{equation}
where $\langle \Sigma \rangle$ denotes the mean entropy production. 
\\
In the case of qudits with dimension greater than two, we have previously shown \cite{ergo} that certain regimes of temperature and frequency allow for ergotropy extraction via unitary operations other than the swap, accompanied by an extended operational range relative to the qubit case. Specifically for qutrits, we provided a full classification of ergotropy-extracting unitaries and demonstrated systematic violations of several known TURs.
\\
In this work, we show that two-mode bosonic heat engines can exhibit analogous enhancements in both SNR and operational range when the swap interaction \cite{max} is replaced by suitably engineered nonlinear transformations. We recall that the swap unitary performs a bilinear mixing of the interacting modes, i.e., when the modes feature different frequencies, it realizes a frequency mixing. Here, we investigate the most general frequency-conversion (FC) interaction between two harmonic oscillators $A$ and $B$ with frequency $\omega_A$ and $\omega_B$, described by the annihilation operators $a$ and $b$, and the corresponding creation operators $a^{\dagger}$ and $b^{\dagger}$, with $[a,a^{\dagger}]=[b,b^{\dagger}]=1$. The general FC unitary reads
\begin{equation} \label{fc}
    V_{\theta} = \exp{(\theta a^{\dagger n}b^m - \theta^* a^nb^{\dagger m})}
\end{equation}
with $n,m\in \mathbb{N}$ and $\theta \in \mathbb{C}$. The choice $n=m=1$ identifies the swap gate mentioned above. Due to its structure, we define the interaction in Eq.~(\ref{fc}) as \textit{polynomial coupling}. Interestingly, the same coupling with $n=1$ and energy-preserving constraint has been recently studied in Ref.~\cite{andolina2024} for the charging of a quantum battery $A$ by the charger $B$, thus achieving the so-called genuine quantum advantage \cite{campaioli2024} with respect to the case $n=m=1$. This result suggests that the polynomial coupling could enhance the engine performance compared to the swap gate. We remark that polynomial couplings and the swap gate correspond to fundamentally different evolution maps: the swap gate preserves Gaussianity, while the polynomial coupling does not.
\\
It is worth noting that the Hamiltonian generating unitary interactions as in Eq.~(\ref{fc}) typically corresponds to nonlinear processes in quantum optics. This is not surprising: the connection with quantum nonlinear optics has been there since the very first foundational investigations of thermodynamics at the microscopic scale \cite{ramsey1956,scovil1959,geusic1967}. In the present case, the unitary in Eq.~(\ref{fc}) can arise from an interaction Hamiltonian between modes $a$ and $b$, with a third mode at frequency $|n\omega_A - m\omega_B|$. This third mode is treated as a classical undepleted coherent pump under parametric approximation \cite{mollow1967, dariano1999}. We also observe that second harmonic generation, i.e. the cases $n=2, m=1$ or $n=1, m=2$, and its quantum effects have been extensively studied from the experimental \cite{franken1961,huang1992,youn1996,zhu2004,frigerio2018,frigerio2021} and theoretical \cite{armstrong1962,crosignani1972,mandel1982,bajer1999,chesi2019,chesi2023exp,chesi2023} point of view throughout the last sixty years, and may thus provide the simplest non-trivial feasible implementation of our engine.
\\
The paper is structured as follows. In section~\ref{2}, we investigate
the bosonic two-stroke heat engine with the general polynomial
coupling of Eq.~(\ref{fc}) for any $n$ and $m$. Here, in section~\ref{2.1} we detail our methods and conventions, in~\ref{2.2} we identify the regime of operation of the engine and evaluate the efficiency, and in~\ref{2.3} we retrieve the distribution of work $p(W)$ up to the second order in the coupling $\theta$ and the condition on $\theta$ such that $0 \leq p(W) \leq 1$. Hence, in section~\ref{2.4} we derive the first moment of the extracted work and, in~\ref{2.5}, the pertaining RFs. We highlight the cases where these quantities are optimized. In section~\ref{3}, we focus on the cases $n=2$, $m=1$ and $n=1$, $m=2$ and expand the interaction $V_{\theta}$ up to the fourth order in $\theta$. We retrieve the distribution of work in section~\ref{3.1}, the average work in~\ref{3.2} and the SNR in~\ref{3.3}, where we also compare our results with the upper bound imposed by the standard TUR \cite{bar,max}. Finally, in section~\ref{4}, we draw our conclusions.

\section{Bosonic two-stroke heat engine with polynomial coupling} \label{2}
\subsection{Framework} \label{2.1}
We start with the most general polynomial coupling, namely $n$ and $m$ in Eq.~(\ref{fc}) can be any positive integer. We use the two-point measurement scheme \cite{deroeck2004,talkner2007cf,talkner2007,esposito2009,campisi2011,ergo}, which allows to estimate simultaneously both work and heat and to derive the joint characteristic function, providing all the moments of these thermodynamic variables. In the following, we fix natural units $\hbar = k_B = 1$.
\\
Each cycle starts with the two harmonic oscillators $A$ and $B$ with Hamiltonians $H_A = \omega_A(a^{\dagger}a+\frac{1}{2})$ and $H_B = \omega_B(b^{\dagger}b+\frac{1}{2})$, at thermal equilibrium with two different pertaining reservoirs, with temperatures $T_A$ and $T_B$ such that $T_A > T_B$. Hence, at the beginning the state of the two modes is given by the tensor product of bosonic Gibbs thermal states, namely
\begin{equation} \label{eq}
    \rho_0 = \frac{e^{-\beta_A H_A}}{Z_A} \otimes \frac{e^{-\beta_B H_B}}{Z_B}
\end{equation}
with the inverse temperature $\beta \equiv T^{-1}$ and the partition function $Z\equiv \Tr[e^{-\beta H}]$. Then the two systems are disconnected from their thermal baths and allowed to interact via the unitary $V_{\theta}$ in Eq.~(\ref{fc}). 
Note that, unlike the swap gate, the interaction for $n > 1$ or $m > 1$ does not preserve the Gaussian character of the initial state $\rho_0$. Indeed, only Hamiltonians that are linear or bilinear in the field modes generate unitary Gaussian operations, which correspond to the so-called metaplectic representation \cite{puri2001,ferraro2005,serafini2023}. As a result, for $n > 1$ or $m > 1$, the interaction $V_{\theta}$ cannot be disentangled in a closed form, and we need to use a Taylor expansion in the coupling parameter $\theta$.
After the interaction, the two modes are reset to the equilibrium state given in Eq.~(\ref{eq}) via full thermalization by their respective baths, thereby completing one cycle of our two-stroke engine. We do not delve into the details of the thermalization process here: since $V_{\theta}$ do not preserve Gaussianity, analyzing it would require solving a master equation for a non-trivial quantum state, which is beyond the scope of this work. However, in the case of the swap gate, this point has been inspected in Ref.~\cite{max2}.
\\
In our description of the adiabatic stroke, the unitary $V_{\theta}$ incorporates all free evolutions, interactions and classical external drivings. Therefore, any energy cost associated with turning the interaction on and off, if present, is inherently accounted for. Indeed, the energy mismatch caused by switching the interaction may itself correspond to the work extracted in each cycle, as discussed in Refs.~\cite{molitor2020, landi2021}. It is worth noting that, while the duration of the unitary stroke can, in principle, be made arbitrary short (for example, by using suitable quenches), the complete thermal relaxation stroke requires a much longer time. Hence, our analysis focuses solely on the work extracted for cycle, rather than on power output.
\\
We characterize the engine by the independent random variables work $W$ extracted by the unitary $V_{\theta}$ and heat $Q_H$ released by the hot reservoir at temperature $T_A$. We fix the convention of positive work when extracted from the systems and positive heat when absorbed from the reservoirs. In each cycle the energy change in system $A$ due to the unitary stroke corresponds to the average heat released by the hot reservoir, namely $\langle Q_H \rangle = - \langle \Delta E_A\rangle$. As well, for the cold reservoir we have $\langle Q_C \rangle = - \langle \Delta E_B\rangle$. Then, the first law of thermodynamics reads $\langle W \rangle = \langle Q_H \rangle + \langle Q_C \rangle$ and the average entropy production reads $\langle \Sigma \rangle = -\beta_A\langle Q_H\rangle -\beta_B\langle Q_C\rangle$. The mean extracted work corresponds to the difference between the mean total energy before and after the interaction stroke, namely
\begin{equation}
    \langle W \rangle \equiv \Tr[\rho_0 (H_A+H_B)] - \Tr[V_{\theta}\rho_0 V_{\theta}^{\dagger} (H_A+H_B)].
\end{equation}

\subsection{Regimes of operation and efficiency} \label{2.2}
The characteristic function obtained from a two-point measurement scheme \cite{talkner2007cf,campisi2011,ergo} for the unitary interaction $V_{\theta}$ reads 
\begin{equation} \label{cf}
\chi(\lambda,\mu) = \Tr[V_{\theta}^{\dagger}e^{-i\mu H_A}e^{-i\lambda(H_A+H_B)}V_{\theta}e^{i\mu H_A}e^{i\lambda(H_A+H_B)}\rho_0]
\end{equation}
where $\lambda$ and $\mu$ identify the work and heat labels, such that all the moments of work and heat can be jointly obtained as follows
\begin{equation} \label{momenta}
\langle W^l Q_H^s \rangle = (-i)^{l+s} \left.\frac{\partial^{l+s}\chi(\lambda,\mu)}{\partial\lambda^l\partial\mu^s}\right |_{\lambda=\mu =0}.
\end{equation}
Since
\begin{equation}
V_{\theta} = e^{i \arg(\theta)a^{\dagger}a/n}V_{|\theta|}e^{-i \arg(\theta)a^{\dagger}a/n},
\end{equation}
it follows from Eq.~(\ref{cf}) that, without loss of generality, the coupling parameter $\theta$ can be assumed as real and positive. This is because the phase of the complex $\theta$ can be ascribed to an irrilevant phase rotation.
By the identity for any bosonic mode $c$
\begin{equation}
          [c,c^{\dagger n}] = nc^{\dagger (n-1)}, \quad \quad  [c^n,c^{\dagger}] = nc^{n-1},
\end{equation}
one can easily check the symmetry
\begin{equation} \label{sym}
[a^{\dagger n}b^m,ma^{\dagger}a+nb^{\dagger}b] = [a^nb^{\dagger m}, ma^{\dagger}a+nb^{\dagger}b] = 0.
\end{equation}
Hence, we can rewrite the characteristic function in Eq.~(\ref{cf}) as
\begin{equation} \label{cf2}
\chi(\lambda,\mu)=\Tr[V_{\theta}^{\dagger}V_{\zeta}\rho_0]
\end{equation}
with $\zeta \equiv \theta e^{-i\xi}$, $\xi \equiv \lambda(n\omega_A - m\omega_B) + \mu n\omega_A$ \cite{max} and $V_\zeta$ as in Eq.~(\ref{fc}), so that
\begin{equation}
\frac{\partial\chi(\lambda,\mu)}{\partial\mu} = \frac{n\omega_A}{n\omega_A-m\omega_B}\frac{\partial\chi(\lambda,\mu)}{\partial\lambda}. \nonumber
\end{equation}
Then, from Eq.~(\ref{momenta}), one obtains
\begin{equation} \label{idmom}
\langle W^lQ_H^s \rangle = \left(\frac{n\omega_A}{n\omega_A - m\omega_B}\right)^s\langle W^{l+s} \rangle.
\end{equation}
This relation together with the first law gives 
\begin{equation}
\langle Q_C \rangle = -\frac{m\omega_B}{n\omega_A}\langle Q_H \rangle,
\end{equation}
and hence the mean entropy production is proportional to the mean work as follows
\begin{equation} \label{mep}
\langle\Sigma\rangle = \frac{m\beta_B\omega_B-n\beta_A\omega_A}{n\omega_A-m\omega_B}\langle W\rangle.
\end{equation}
The second law of thermodynamics, namely the positivity of $\langle \Sigma \rangle$, allows to identify the following three standard regimes of operation
\begin{enumerate}
\item 
\begin{equation}
\frac{nT_B}{mT_A} < \frac{\omega_B}{\omega_A} < \frac{n}{m} \quad\quad\quad {\rm heat \, engine;} \label{he}
\end{equation}
\item
\begin{equation}
\frac{\omega_B}{\omega_A} < \frac{nT_B}{mT_A} \quad\quad\quad\quad\quad {\rm refrigerator;}
\end{equation}
\item
\begin{equation}
\frac{\omega_B}{\omega_A} > \frac{n}{m} \quad\quad\quad\quad\quad {\rm thermal \, accelerator;}
\end{equation}
\end{enumerate}
where, correspondingly, we have
\begin{enumerate}
\item 
\begin{equation}
\langle W \rangle >0, \quad \langle Q_H \rangle > 0, \quad \langle Q_C \rangle < 0;
\end{equation}
\item
\begin{equation}
\langle W \rangle <0, \quad \langle Q_H \rangle < 0, \quad \langle Q_C \rangle > 0;
\end{equation}
\item
\begin{equation}
\langle W \rangle <0, \quad \langle Q_H \rangle > 0, \quad \langle Q_C \rangle < 0.
\end{equation}
\end{enumerate}
Here we focus on the extraction of positive work, i.e. on the heat engine, but all our results can be exploited to assess the performance of refrigerators and accelerators as well. The mean extracted work in the regime (i) is the average energy released by the annihilation of $n$ bosons at frequency $\omega_A$ and the creation of $m$ bosons at frequency $\omega_B < n\omega_A/m$.
\\
We observe that in this scenario both the range of operation and the efficiency of the engine drastically change with respect to the case of the swap gate \cite{max}. In particular, if $n > m$, they both increase. Let us rename the frequency ratio $\omega_B/\omega_A \equiv x$ and the corresponding boundaries of the positive-work regime $x_{\rm min}$ and $x_{\rm max}$. Then we see that the range of operation $\Delta x \equiv x_{\rm max} - x_{\rm min}$ grows linearly with $x_{\rm max} = n/m$, i.e.
\begin{equation}
\Delta x = \eta_Cx_{\rm max}
\end{equation}
where the proportionality is given by the Carnot efficiency $\eta_C \equiv 1 - T_B/T_A$. Therefore, we find an enhancement of a factor  $x_{\rm max}$ with respect to the range of operation where the swap interaction extracts work. The efficiency $\eta \equiv \langle W \rangle / \langle Q_H \rangle$ can be easily retrieved from Eq.~(\ref{idmom}) and reads
\begin{equation}
\eta = 1 - \frac{m\omega_B}{n\omega_A} = 1 - \frac{x}{x_{\rm max}} \leq \eta_C
\end{equation}
with equality for $x = x_{\rm min}$. 

\subsection{Distribution of work and heat} \label{2.3}
\begin{figure}
\centering
\includegraphics{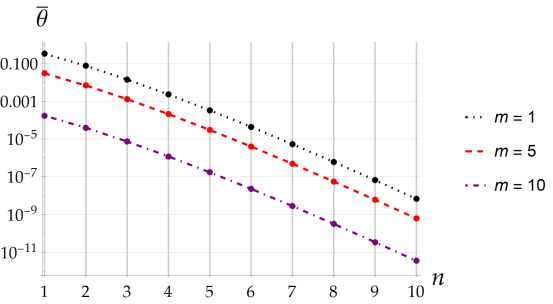}
\caption{Upper bound $\bar{\theta}$ in Eq.~(\ref{uptheta}) on the coupling factor $\theta$ as a function of $n$, with $\beta_A\omega_A = 0.1$ and $\beta_B\omega_B = 10$.}
\label{figtheta}
\end{figure}

The joint probability distribution of work and heat $p(W,Q_H)$ is the inverse Fourier transform of the characteristic function $\chi(\lambda,\mu)$ in Eq.~(\ref{cf}). From Eq.~(\ref{cf2}) one finds that $\chi(\lambda, \mu)$ is periodic in $\lambda$ and $\mu$, with periods $\Lambda \equiv 2\pi/(n\omega_A-m\omega_B)$ and $M \equiv 2\pi/n\omega_A$. Hence, the joint distribution $p(W, Q_H)$ is discrete, with $W$ and $Q_H$ as integer multiples of $n\omega_A - m\omega_B$ and $n\omega_A$, namely
$$
p[W = j(n\omega_A - m\omega_B), Q_H = kn\omega_A] = \frac{1}{\Lambda M}\int_{-\Lambda/2}^{\Lambda/2}d\lambda\int_{-M/2}^{M/2}d\mu\,\chi(\lambda,\mu) e^{-ij(n\omega_A - m\omega_B)\lambda - ikn\omega_A\mu}.
$$
Since $\chi(\lambda,\mu)$ is actually a function of the single variable $\xi \equiv  \lambda(n\omega_A - m\omega_B) + \mu n\omega_A$ one has
\begin{equation} \label{pcf}
p[W = j(n\omega_A - m\omega_B), Q_H = kn\omega_A] = \delta_{j,k}\, \frac{1}{2\pi}\int_0^{2\pi}d\mu\, \chi(0,\frac{\mu}{n\omega_A})e^{-ik\mu}
\end{equation}
which shows that the random variables $W$ and $Q_H$ are perfectly
correlated. In the following, since the algebra of $\{a^{\dagger n}b^m, a^nb^{\dagger m}\}$ is not closed for $n \neq 1$ and $m \neq 1$, in order to evaluate $\chi(\lambda,\mu)$, we expand $V_{\theta}$ up to the second order in $\theta$. Given the Gibbs state $R_{\beta_C}$ for a bosonic mode $c$
\begin{equation}
R_{\beta_C} = \frac{e^{-\beta_C\omega_Cc^{\dagger}c}}{\Tr[e^{-\beta_C\omega_Cc^{\dagger}c}]} = \frac{1}{N_C+1}\sum_{l=0}^{\infty}\left(\frac{N_C}{N_C+1}\right)^l|l\rangle\langle l|,
\end{equation}
where $N_C \equiv (e^{\beta_C\omega_C}-1)^{-1}$ denotes the average population, we can exploit the following identities:
\begin{eqnarray}
\Tr[c^nc^{\dagger m}R_{\beta_C}] = \delta_{n,m}\frac{1}{N_C+1}\sum_{l=0}^{\infty}\left(\frac{N_C}{N_C+1}\right)^l\frac{(l+n)!}{l!} = n!(N_C+1)^n \label{id1} \\
\Tr[c^{\dagger n}c^mR_{\beta_C}] = \delta_{n,m}\frac{1}{N_C+1}\sum_{l=0}^{\infty}\left(\frac{N_C}{N_C+1}\right)^l\frac{l!}{(l-n)!} = n!N_C^n. \label{id2}
\end{eqnarray}
By making use of Eqs.~(\ref{id1}) and~(\ref{id2}), we can evaluate the correlation function $\chi(\lambda, \mu)$ from Eq.~(\ref{cf2}). Up to the second order, we find
\begin{eqnarray}
\chi(\lambda, \mu) = & 1+ \theta^2n!m!\{[(N_A+1)^nN_B^m+N_A^n(N_B+1)^m][\cos(\xi)-1] \nonumber \\
& +\, i[(N_A+1)^nN_B^m-N_A^n(N_B+1)^m]\sin(\xi)\} + O(\theta^4). \label{cfnl}
\end{eqnarray}
Then, by Eq.~(\ref{pcf}), the joint probability distribution of $W$ and $Q_H$ up to the order $\theta^2$ is the following 3-point distribution
\begin{eqnarray}
p(W,Q_H) &\simeq\delta(W)\delta(Q_H)\{1-\theta^2n!m![(N_A+1)^nN_B^m+N_A^n(N_B+1)^m]\} \nonumber \\
&+ \delta(W+n\omega_A-m\omega_B)\delta(Q_H+n\omega_A)\theta^2n!m!(N_A+1)^nN_B^m \nonumber \\
&+ \delta(W-n\omega_A+m\omega_B)\delta(Q_H-n\omega_A)\theta^2n!m!N_A^n(N_B+1)^m.
\end{eqnarray}
In the regime of heat-engine operation of Eq.~(\ref{he}), notice that the probability of obtaining $W = n\omega_A-m\omega_B > 0$ is strictly larger than the one of having $W = - (n\omega_A-m\omega_B) < 0$ due to the condition $n\beta_A\omega_A < m\beta_B\omega_B$. In fact, the ratio between emission and absorption is given by 
\begin{equation}
\frac{p(W)}{p(-W)} = \left(\frac{N_A}{N_A + 1}\right)^n\left(\frac{N_B+1}{N_A}\right)^m = e^{m\beta_B\omega_B - n\beta_A\omega_A}, \nonumber
\end{equation}
which competes with the amount of energy of the quantum $n\omega_A - m\omega_B$. However, the emission/absorption ratio can be usefully maximized by taking extreme temperatures $\beta_A \rightarrow 0$, $\beta_B \rightarrow \infty$ and keeping the frequency ratio strictly within the regime of operation, i.e. $\omega_B/\omega_A < n/m$. This maximization is doable as long as the coupling factor $\theta$ is correspondingly tuned.
The positivity and the normalization of $p(W,Q_H)$ yield an upper bound on $\theta$, necessary for the expansion to make sense. The condition $p(W=0,Q_H=0)\geq 0$ provides the strictest bound on $\theta$, which is given by
\begin{equation} \label{uptheta}
\theta \leq \{n!m![(N_A+1)^nN_B^m+N_A^n(N_B+1)^m]\}^{-1/2} \equiv \bar{\theta}.
\end{equation}
In Fig.~\ref{figtheta}, we show some extreme values of $\theta$ for $n,m\in [1,10]$, with $\beta_A\omega_A = 0.1$ and $\beta_B\omega_B = 10$. It is worth noting that, whenever $\beta\omega \ll 1$, the classical equipartition of energy is recovered, since $N \equiv (e^{\beta\omega} - 1)^{-1} \simeq (\beta\omega)^{-1}$, thus yielding a purely thermal energy $\epsilon = \omega N \simeq T$.

\subsection{Mean extracted work} \label{2.4}
\begin{figure}
\includegraphics[scale=0.23]{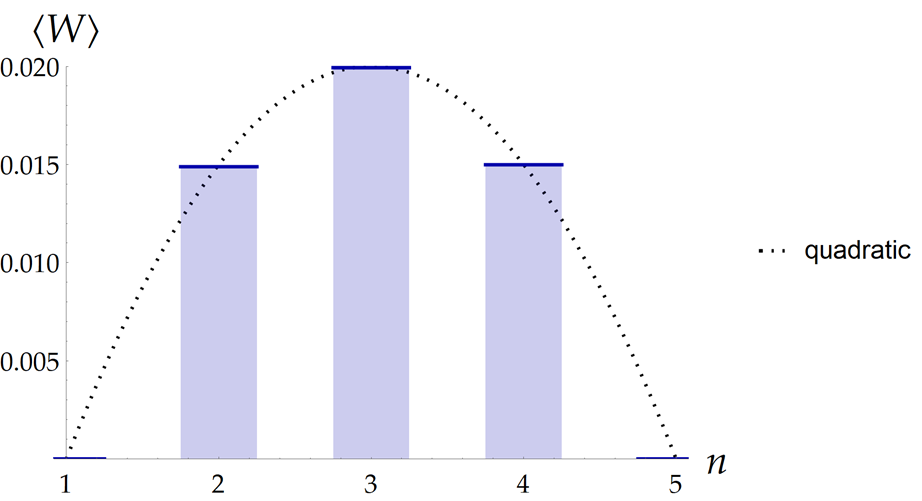} \hspace{0.5cm}
\includegraphics[scale=0.23]{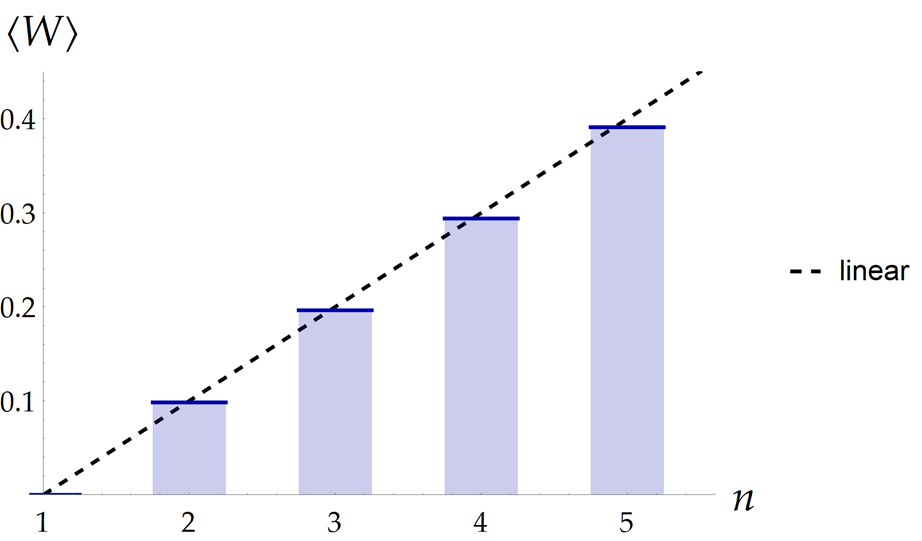}
\caption{The histograms display the mean extracted work [see Eq.~(\ref{mw2})] as a function of $n$, with $\alpha = 0.1$, $\omega_A = 1$, $\omega_B = 0.5$, $\beta_A = 0.1$, $m=2$. Left: $\beta_B = 0.5$. The dotted black curve displays the parabolic approximation to Eq.~(\ref{mw2}). Right: $\beta_B = 5$. The dashed black line displays the linear approximation to Eq.~(\ref{mw2}).}
\label{appmw}
\end{figure}

Now we can evaluate all the moments of the extracted work up to the second order in $\theta$. The mean value reads
\begin{equation}
\langle W \rangle = \theta^2(n\omega_A-m\omega_B)n!m![N_A^n(N_B+1)^m - (N_A+1)^nN_B^m] + O(\theta^4). \label{mw}
\end{equation}
We notice that $\langle W \rangle$ can be optimized over the choice of the polynomial coupling, i.e. over $n$ and $m$, for fixed frequencies and temperatures. In fact, $\langle W \rangle > 0$ for 
\begin{equation} \label{posW}
\frac{\omega_B}{\omega_A} < \frac{n}{m} < \frac{\beta_B\omega_B}{\beta_A\omega_A}.
\end{equation}
However, a fair optimization needs to take into account the upper bound $\bar{\theta}$ on the coupling constant $\theta$ in Eq.~(\ref{uptheta}). This parameter strongly depends on the details of the nonlinear process. Let us suppose that one can achieve a fraction $\alpha$ of the upper bound $\bar{\theta}$, i.e. we set $\theta = \sqrt{\alpha} \bar{\theta}$ and keep $\alpha$ small, i.e. $\alpha \in (0,0.5]$, since couplings that typically feature nonlinear processes are weak. Hence, the probability that the engine does not extract nor absorb work is fixed to $p(W=0) = 1-\alpha$. This assumption greatly simplifies the optimization of the average work, since Eq.~(\ref{mw}) can be re-expressed as
\begin{equation}
\langle W \rangle = \alpha(n\omega_A-m\omega_B)\,{\rm tgh}\left(\frac{m\beta_B\omega_B - n\beta_A\omega_A}{2}\right)+ O(\theta^4) \label{mw2}
\end{equation}
Note that in general $\langle W \rangle < \alpha(n\omega_A-m\omega_B)$, since ${\rm tgh}(x) < 1$ $\forall\,x > 0$. The expression in Eq.~(\ref{mw2}) further simplifies in the two extreme regimes where $m\beta_B\omega_B \simeq n\beta_A\omega_A$ and $m\beta_B\omega_B \gg n\beta_A\omega_A$. In the former case, we find $\langle W \rangle \simeq \alpha(n\omega_A-m\omega_B)(m\beta_B\omega_B - n\beta_A\omega_A)/2$. In the latter, $\langle W \rangle$ is almost linear with $n\omega_A-m\omega_B$, which means that it hardly depends on the temperatures. In Fig.~\ref{appmw}, we highlight these two different settings by plotting $\langle W \rangle$ as a function of $n$, with fixed $m = 2$. On the left, the chosen parameters let the hyperbolic tangent be large enough to yield a non-negligible contribution for every choice of $n$. Therefore, $\langle W \rangle$ never approaches the linear behavior. On the right, the temperature $T_B$ of the cold reservoir is one order of magnitude smaller, $\langle W \rangle$ is linear versus $n$, and hence is maximized by the largest $n$ experimentally achievable for fixed $m$.

\begin{figure}
\includegraphics[scale=0.25]{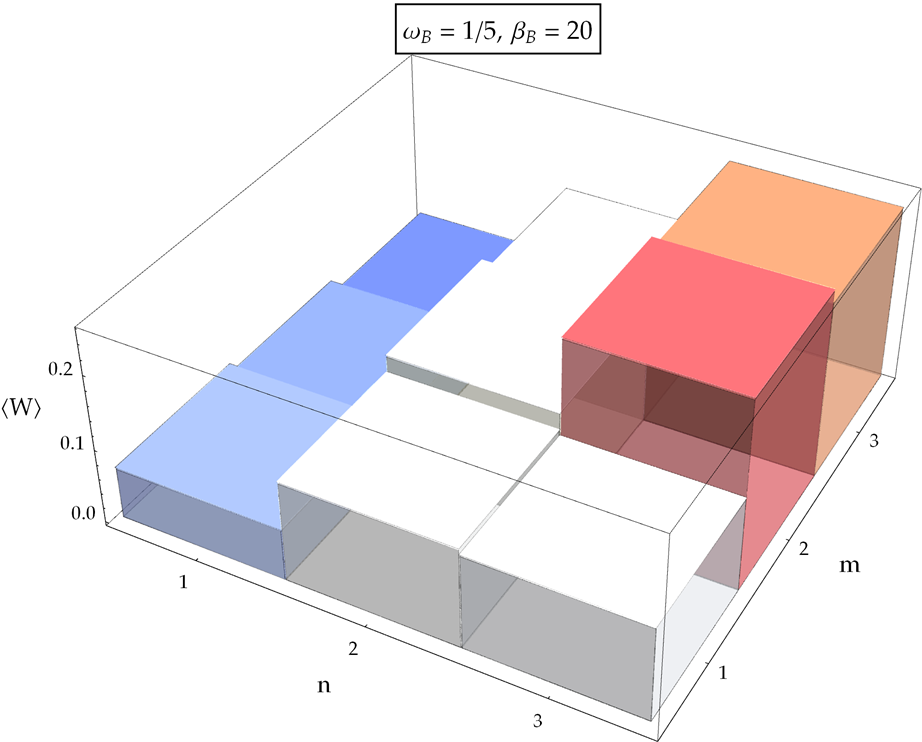} \hspace{0.25cm}
\includegraphics[scale=0.25]{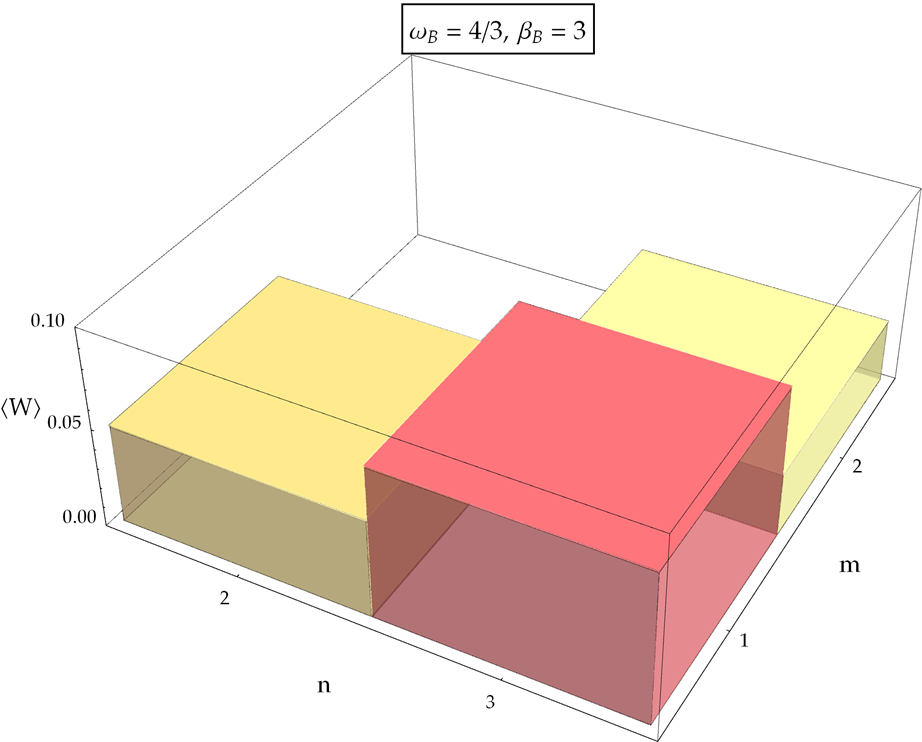}
\caption{Mean extracted work $\langle W \rangle$ in Eq.~(\ref{mw3}) as a function of $n$ and $m$ with $\alpha = 0.1$, $\beta_A = 1$, $\omega_A=1$.Left: $\omega_B=1/5$, $\beta_B=20$. Right: $\omega_B=4/3$, $\beta_B=3$.}
\label{xy}
\end{figure}

It is simple to provide a numeric estimation of the maximum of $\langle W \rangle$ versus $n$ and $m$ for specific settings. 
To this aim, we fix the ratios of the inverse temperatures $y\equiv \beta_B/\beta_A$ and of the frequencies $x \equiv \omega_B/\omega_A$ and rewrite Eq.~(\ref{mw2}) as
\begin{equation}
\langle W \rangle = \alpha\omega_A(n-mx)\,{\rm tgh}\left(\frac{\beta_A\omega_A(mxy - n)}{2}\right)+ O(\theta^4) \label{mw3}.
\end{equation}
Hence, if $\beta_A\omega_A \gg 1$, the mean work is described by the linear regime for every $n$ and $m$ with $n/m < xy$.
If this is not the case and $\beta_A\omega_A > (mxy - n)/2$ for some $n$ and $m$, then there is a competition between the factor $n\omega_A - m\omega_B$ and the temperature-dependent contribution of the hyperbolic tangent. We address this in Fig.~\ref{xy}, where we set $\beta_A = \omega_A = 1$.
In particular, we keep fixed the upper bound of $x_{\rm max} = n/m$, namely $(\beta_B\omega_B)/(\beta_A\omega_A) = xy = 4$ and focus on the cases $x < 1$, i.e. $\omega_B < \omega_A$, (left panel) and $x > 1$, i.e. $\omega_B > \omega_A$, (right panel). The competition between the two contributions is clear in the first regime, where we note that the optimality is achieved for $1< m \leq n $. Conversely, in the second case, from Eq.~(\ref{posW}) the condition $\omega_B > \omega_A$ implies $n > m$, and we observe in the right panel that the maximum work is obtained for $m=1$ and $n = xy -1$. We note also that in this regime the swap gate does not extract work, and we need $n \geq 2$.

\begin{figure}
\includegraphics[scale=0.53]{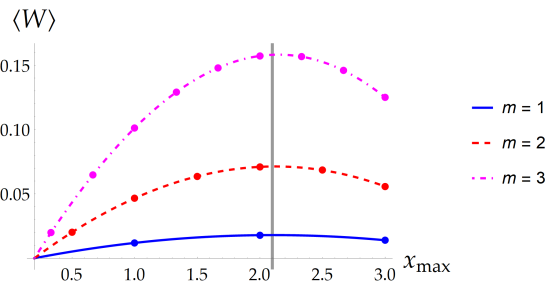} \hspace{0.5cm}
\includegraphics[scale=0.53]{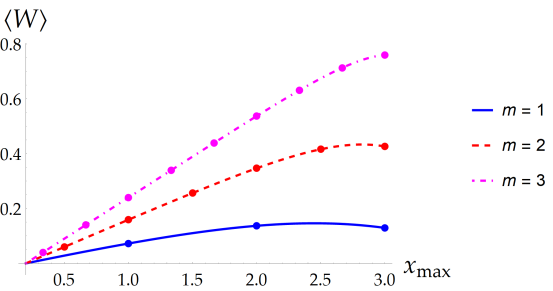}
\caption{Mean extracted work $\langle W \rangle$ in Eq.~(\ref{mw3}) as a function of $x_{\rm max} = n/m$ with $\alpha = 0.1$,  $\omega_A = 1$, $\beta_B/\beta_A = 20$, $\omega_B=1/5$. Left: $\beta_A = 0.1$. Right: $\beta_A = 1$. The grey vertical line in the left panel identifies the approximated maximum of $\langle W \rangle$ at $x_{\rm max}^{\rm (opt)}$ in Eq.~(\ref{xopt1}).}
\label{anmaxw}
\end{figure}

If $\beta_A\omega_A < (mxy - n)/2$ $\forall\, n, m$ such that $x < x_{\rm max} < xy$, then $\langle W \rangle$ approaches the quadratic regime, where it is maximum for
\begin{equation} \label{xopt1}
x_{\rm max}^{\rm (opt)} \simeq \frac{\omega_B(\beta_B+\beta_A)}{2\beta_A\omega_A} = x\,\frac{y+1}{2}.
\end{equation}
We show the dependence of $\langle W \rangle$ on $x_{\rm max}$ in the panels of Fig.~\ref{anmaxw}, where two regimes are considered: in the left panel we set $\beta_A\omega_A = 0.1$ and the behavior of the mean work is distinctly quadratic, while this is not the case for the right panel, where we set $\beta_A\omega_A = 1$. In the first panel the maximum of $\langle W \rangle$ is approximately identified by the optimal  $x_{\rm max}$ (grey vertical line) in Eq.~(\ref{xopt1}). The curves display three values of $m$ and the points show the ratios $x_{\rm max}$ for the corresponding values of $n$. The other settings are fixed as in the left panel of Fig.~\ref{xy}. It is worth noting that in the case $\beta_A\omega_A = 0.1$, where for system $A$ the classical energy equipartition holds, the mean work extracted by the swap gate is outperformed by all the polynomial interactions considered.

\subsection{Work fluctuations} \label{2.5}
\begin{figure}
\includegraphics[scale=0.25]{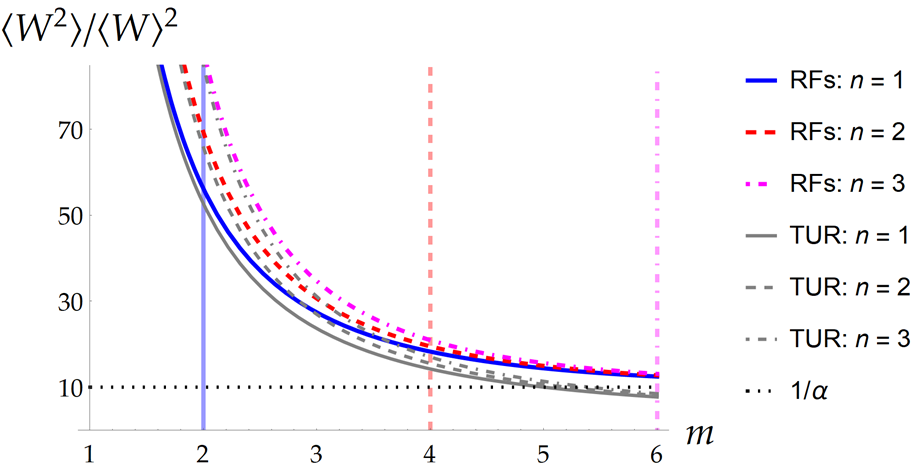} \hspace{0.25cm}
\includegraphics[scale=0.25]{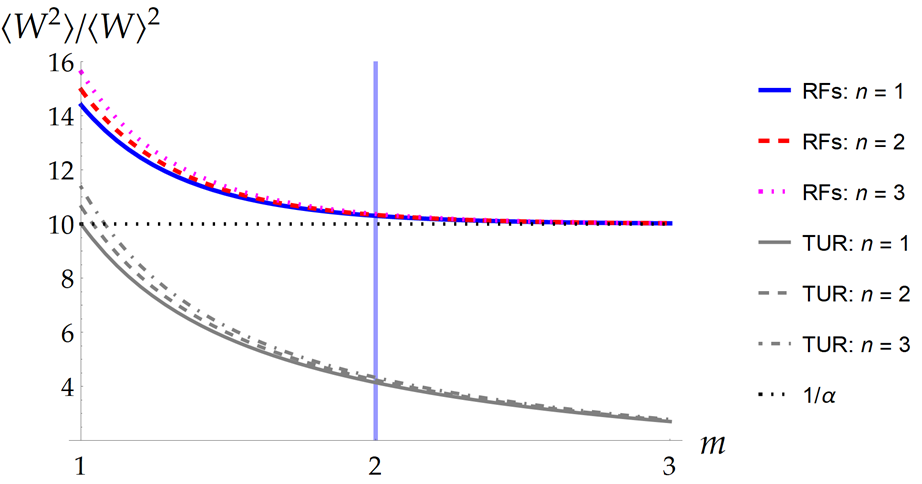}
\caption{RFs of the extracted work in Eq.~(\ref{rfmin}) as a function of $m$ for $n = 1$ (solid blue curve), $n = 2$ (dashed red curve) and $n = 3$ (dot-dashed magenta curve) with $\alpha = 0.1$,  $\omega_A = 1$, $\omega_B = 0.5$, $\beta_A = 0.1$. The asymptotic limit $\alpha^{-1}$ obtained for large $m$ is displayed as a black dotted line. The vertical lines identify the largest $m$ compatible with the extraction of positive mean work, namely $m = 2n$ with $n = 1$ (solid blue line), $n = 2$ (dashed red line) and $n = 3$ (dot-dashed magenta line). The gray thin curves show the lower bounds given by the TUR in Eq.~(\ref{tur}), for $n = 1$ (solid), $n = 2$ (dashed) and $n = 3$ (dot-dashed),  Left: $\beta_B = 1$. Right: $\beta_B = 5$.}
\label{rf}
\end{figure}

The second moment of the extracted work is given by
\begin{eqnarray} \label{smw}
\langle W^2 \rangle = &\theta^2(n\omega_A-m\omega_B)^2n!m![N_A^n(N_B+1)^m + (N_A+1)^nN_B^m] \\
& + O(\theta^4). \nonumber
\end{eqnarray}
We observe that at the order $\theta^2$ the variance equals the second moment, i.e. ${\rm var}(W) = \langle W^2\rangle$. Therefore, we can straightforwardly evaluate the RFs and relate them with the mean entropy production, retrieved in Eq.~(\ref{mep}). We find
\begin{eqnarray}
\frac{{\rm var}(W)}{\langle W \rangle^2} &= \frac{{\rm var}(Q_H)}{\langle Q_H \rangle^2} = \frac{{\rm cov}(W, Q_H)}{\langle W \rangle\langle Q_H \rangle}  \nonumber \\
&\simeq \frac{m\beta_B\omega_B - n\beta_A\omega_A}{\langle \Sigma \rangle}\frac{N_A^n(N_B+1)^m + (N_A+1)^nN_B^m}{N_A^n(N_B+1)^m - (N_A+1)^nN_B^m} \nonumber \\
&= \frac{h(m\beta_B\omega_B - n\beta_A\omega_A)}{\langle \Sigma \rangle} \label{rf2ord}
\end{eqnarray}
where we introduced the function $h(z) \equiv z\,{\rm cotgh}(z/2)$. Since $h(z) \geq 2$, the standard TUR \cite{bar}
\begin{equation} \label{tur}
\frac{{\rm var}(W)}{\langle W \rangle^2} \geq \frac{2}{\langle \Sigma \rangle}
\end{equation}
is never violated. However, the TUR presented in Eq.~(\ref{tur}) is less stringent than the bound in Eq.~(\ref{maxtur}), which holds in the case $n=1$ and $m=1$ as shown in Ref.~\cite{max}. This relaxation stems from the truncation of the interaction $V_{\theta}$ at second order in the coupling parameter $\theta$, an approximation that remains valid and physically realistic when fourth-order contributions are negligible. Notably, the lower bound in Eq.~(\ref{tur})  depends on $\theta^2$ through the mean entropy production, highlighting the sensitivity of the resulting TUR to the perturbative order considered in the expansion of  $V_{\theta}$. The influence of the truncation order on the implied TURs is significant. In particular, as we will show in section~\ref{3.3}, expanding $V_{\theta}$ to fourth order further relaxes the TURs for the cases $n=2,\, m=1$, and $n=1,\, m=2$. The validity of the TUR in Eq.~(\ref{tur}) within this perturbative framework is ensured by the upper bound on $\theta$ given in Eq.~(\ref{uptheta}), which preserves the consistency of the expansion.
\\
From Eq.~(\ref{uptheta}), we can write
\begin{equation}
\langle W^2 \rangle \simeq \alpha(n\omega_A-m\omega_B)^2,
\end{equation}
and hence, by Eq.~(\ref{mw2}), one obtains
\begin{eqnarray} \label{rfmin}
\frac{{\rm var}(W)}{\langle W \rangle^2} &\simeq \frac{1}{\alpha}{\rm cotgh}^2\left(\frac{m\beta_B\omega_B - n\beta_A\omega_A}{2}\right) \\
&\geq \frac{1}{\alpha}{\rm cotgh}^2\left[\frac{m\omega_B}{2}(\beta_B - \beta_A)\right] \nonumber
\end{eqnarray}
For fixed frequencies and temperatures, we notice that the RFs are lowered by setting $m \gg n =1$, as long as $m < n\omega_A/\omega_B$, in order to have positive work extraction. This is easily seen in Fig.~\ref{rf}, where we 
fixed $n = 1$ (blue solid curve), $n = 2$ (red dashed curve) and $n =
3$ (magenta dot-dashed curve) and plotted the RFs as a function of
$m$. In the setting of Fig.~\ref{rf}, the condition for the extraction
of positive mean work reads $m < 2n$ and is identified for each choice
of $n$ as a vertical line. The two panels feature a different
temperature of the cold reservoir: in the left panel $\beta_B$ is
smaller, implying larger RFs at small values of $m$. For large $m$,
the lower bound in Eq.~(\ref{rfmin}) tends to $\alpha^{-1}$, which is
marked as a black dotted horizontal line. From both panels, we find a
competition between the extraction of positive work and the
minimization of the RFs: the latter requires a small $n$, which turns
into a restriction on $m$. This competition hampers  the performance
of the engine in the case of the left panel, where we have $\beta_B\omega_B \gtrsim \beta_A\omega_A$, for which we optimize $\langle W \rangle$ if $m \simeq n$: here, for $n = 1$ and $n = 2$, we do not even achieve minimal fluctuations in the regime where $\langle W \rangle > 0$. Instead, in the case $\beta_B\omega_B \gg \beta_A\omega_A$, on the right, we can optimize $\langle W \rangle$ by selecting $n > m$, while achieving near optimal RFs. We also show (gray lines) the lower bound from the standard TUR in Eq.~(\ref{tur}) for each choice of $n$.

\section{Work extraction from Second Harmonic Generation} \label{3}
Now we focus on the nonlinear processes where two bosons of one mode are converted into one boson of the other mode, namely we consider Eq.~(\ref{fc}) with $n = 2$ and $m = 1$, or $n = 1$ and $m = 2$. We will refer shortly to the former as $V_{21}$ and to the latter as $V_{12}$.
\\
Here, we fix the ratio of the inverse temperatures as follows: $\beta_B / \beta_A \equiv y = 10^2$. Moreover, in the case of $V_{21}$, we fix the ratios of the frequencies to the values $\omega_B/\omega_A \equiv x \in \{1/10,1/3,1/2,2/3,1,3/2\}$, while, in the case of $V_{12}$, to $x \in \{1/150,1/100,1/10,1/8,1/4,1/3\}$. Within these choices, according to Eq.~(\ref{he}), we are in the heat-engine regime, and we have $\beta_B\omega_B = xy\beta_A\omega_A$. For this case, we will expand the nonlinear interactions up to the fourth order in the coupling $\theta$.

\subsection{Distribution of work} \label{3.1}
\begin{figure}
\includegraphics[scale=0.23]{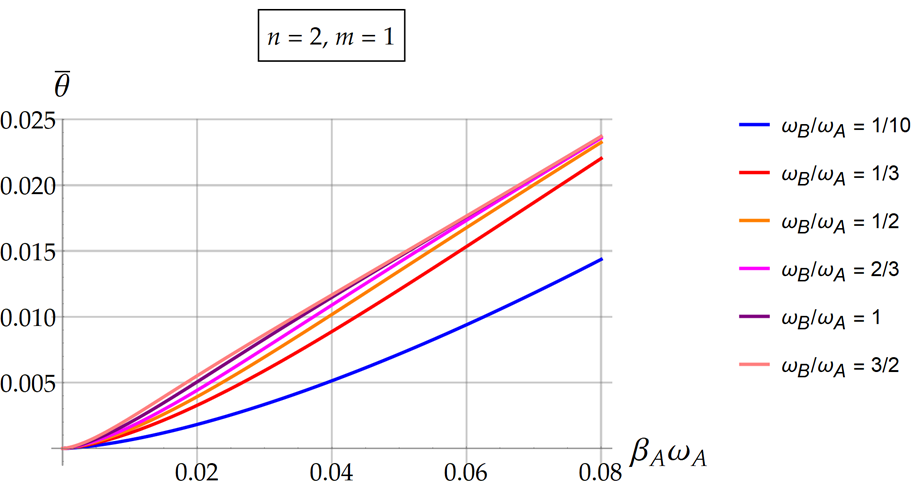} \hspace{0.5cm}
\includegraphics[scale=0.23]{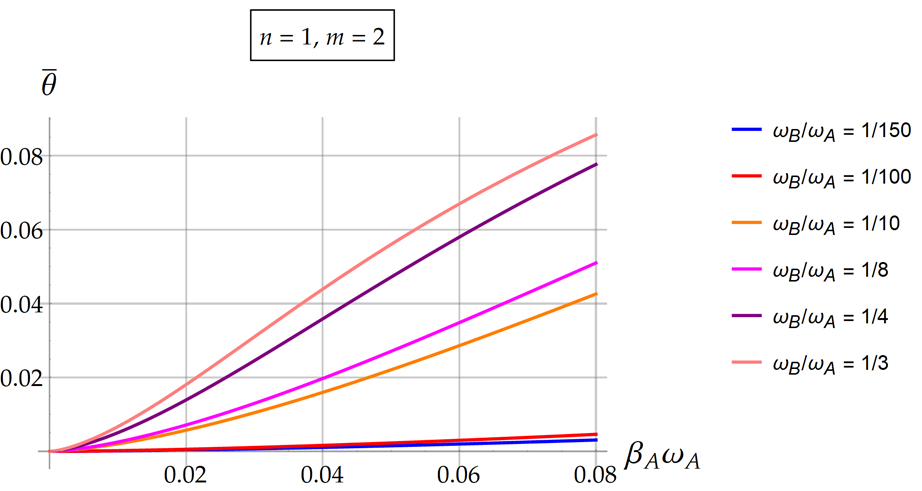}
\caption{Upper bound on the coupling $\theta$ that guarantees $0 \leq p[W = k(2\omega_A - \omega_B)] \leq 1$ for $k = 0, \pm 1, \pm 2$. Here we set $\beta_B/\beta_A = 10^2$ and show the maximal couplings $\bar{\theta}$ as a function of $\beta_A\omega_A$ for different choices of $\omega_B/\omega_A$.}
\label{thetamaxshg}
\end{figure}

The stochastic work $W$ is distributed according to a 5-point expression where $W = k(n\omega_A - m\omega_B)$, with $k = 0, \pm 1, \pm 2$. In the case of $V_{21}$, one has explicitly
\begin{eqnarray}
&p(W = 0) \simeq\delta(W)[1-2\theta^2\mathcal{A} + 4\theta^4(\mathcal{A}\mathcal{B}-\mathcal{C})] \label{one}  \\
&p(W = 2\omega_A - \omega_B) \simeq \delta(W-2\omega_A+\omega_B)2N_A^2(N_B+1)(\theta^2 - 2\theta^4\mathcal{B}) \label{p-} \\
&p(W = \omega_B - 2\omega_A) \simeq \delta(W+2\omega_A-\omega_B)2(N_A + 1)^2N_B(\theta^2 - 2\theta^4\mathcal{B}) \label{p+} \\
&p(W = 2(2\omega_A - \omega_B)) \simeq \delta(W-2(2\omega_A-\omega_B))12\theta^4N_A^4(N_B+1)^2 \\
&p(W = 2(\omega_B - 2\omega_A)) \simeq \delta(W+2(2\omega_A-\omega_B))12\theta^4(N_A+1)^4N_B^2,
\end{eqnarray}
where
\begin{eqnarray} \label{A}
&\mathcal{A} = N_A^2(N_B+1) + (N_A+1)^2N_B \\
&\mathcal{B} = (2N_B+1)(6N_A^2 + 6N_A + 1) - \frac{2}{3}\left(6N_A - 4N_B + 1\right) \label{B} \\
&\mathcal{C} = 3[N_A^4(N_B + 1)^2 + (N_A+1)^4N_B^2]. \label{C}
\end{eqnarray}
Notice that the bound $n\beta_A\omega_A > m\beta_B\omega_B$ for heat-engine operation, in the present scenario, is equivalent to $N_A^2(N_B+1) > (N_A + 1)^2N_B$.
In the case of $V_{12}$, $p(W)$ is obtained by exchanging $N_A$ with $N_B$. We remark that $\langle W \rangle > 0$ for $V_{21}$ and $V_{12}$ if $\omega_B < 2\omega_A$ and $\omega_A > 2\omega_B$, respectively. In this regime, the quanta of work $\varepsilon_{21} \equiv 2\omega_A - \omega_B$ and $2\varepsilon_{21}$ are \textit{extracted} by $V_{21}$, while $-\varepsilon_{21}$ and $-2\varepsilon_{21}$ are \textit{absorbed}. The same analysis holds for $V_{12}$ with $\varepsilon_{12} = \omega_A - 2\omega_B$.
\\
At the fourth order in $\theta$, the tightest requirement on the coupling is given by the positivity of the probabilities in Eqs.~(\ref{p-}) and~(\ref{p+}), yielding
\begin{equation} \label{thetaup2}
\theta \leq (2\mathcal{B})^{-\frac{1}{2}} \equiv \bar{\theta}.
\end{equation}
This condition also implies $0 < p[W = k(2\omega_A - \omega_B)] < 1$ for $k = 0, \pm 2$. We show the upper bounds on the coupling in Fig.~\ref{thetamaxshg} for $V_{21}$ (left panel) and $V_{12}$ (right panel), as a function of $\beta_A\omega_A$, with $\beta_B\omega_B = xy\beta_A\omega_A$. 

\begin{figure}
\includegraphics[scale=0.2]{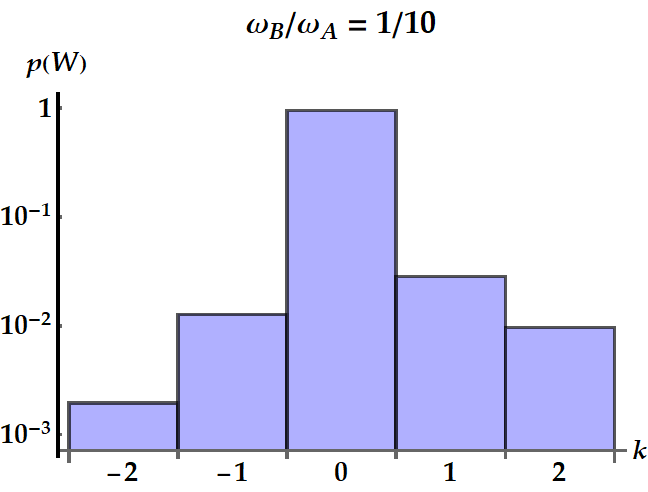} \hspace{0.2cm}
\includegraphics[scale=0.2]{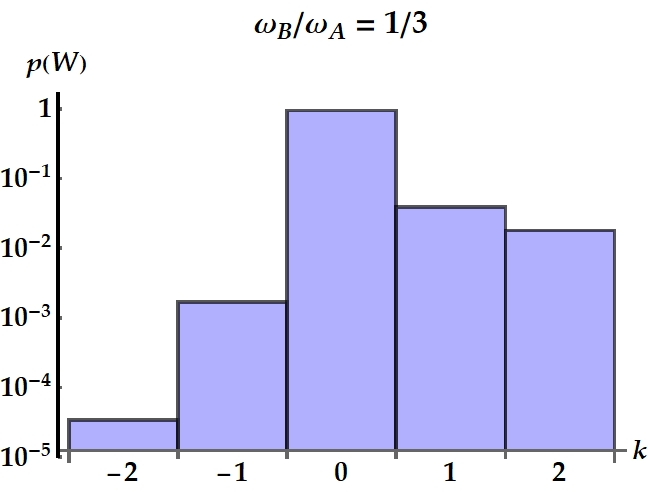} \hspace{0.2cm}
\includegraphics[scale=0.2]{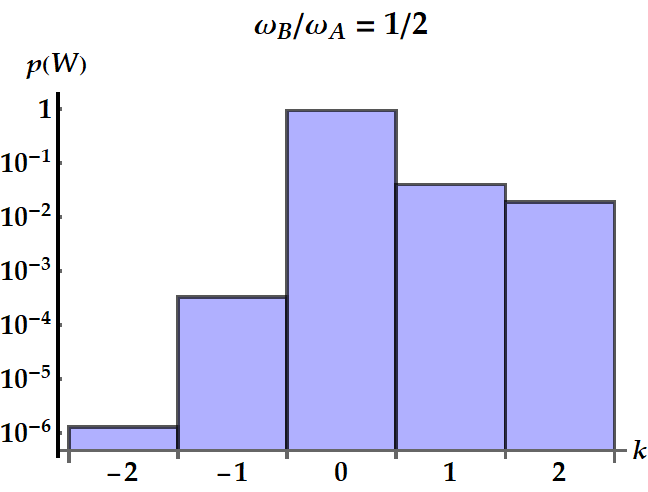} \\ \\
\includegraphics[scale=0.2]{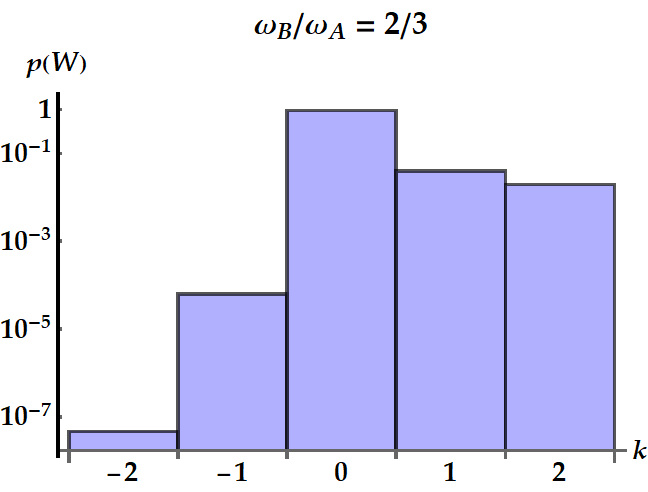} \hspace{0.2cm}
\includegraphics[scale=0.2]{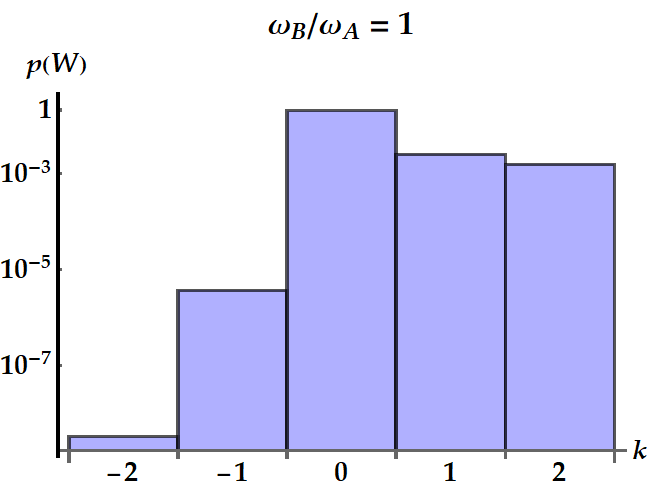} \hspace{0.2cm}
\includegraphics[scale=0.2]{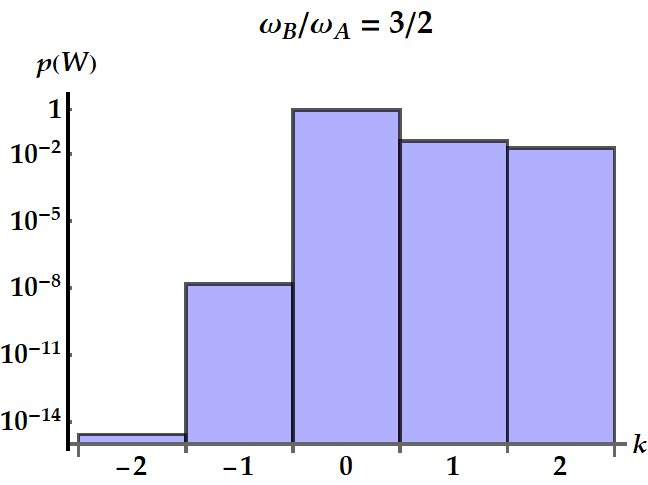}
\caption{Logarithmic plot of the distribution of work $p(W)$ given by the expansion up to the fourth order of $V_{21}$, with $\alpha = 0.5$,  $\beta_B/\beta_A=10^2$, $\beta_A\omega_A = 0.1$.}
\label{p}
\end{figure}

\begin{figure}
\includegraphics[scale=0.2]{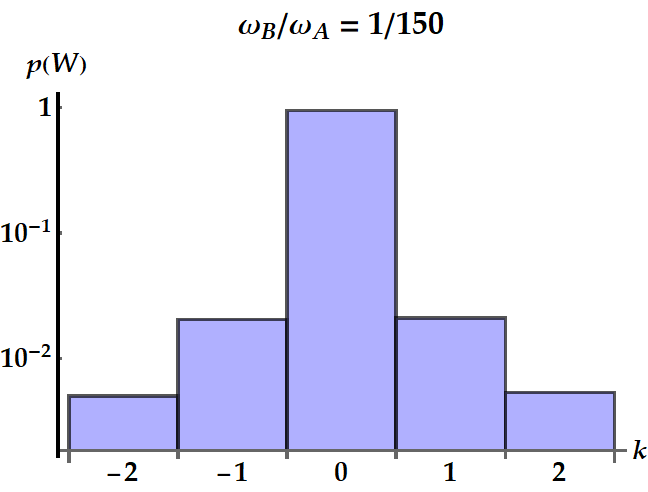} \hspace{0.2cm}
\includegraphics[scale=0.2]{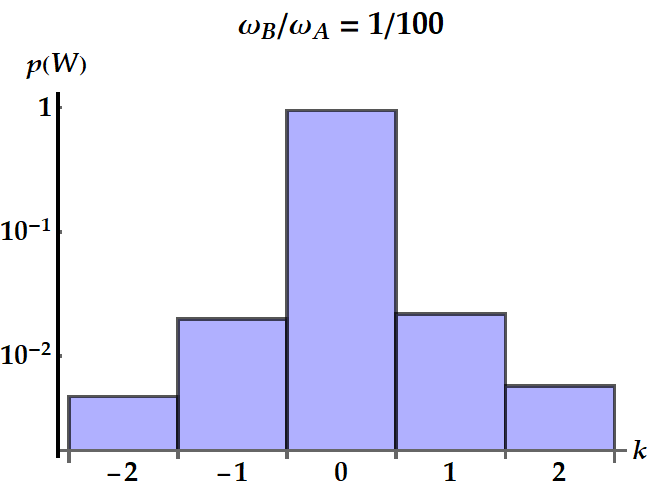} \hspace{0.2cm}
\includegraphics[scale=0.2]{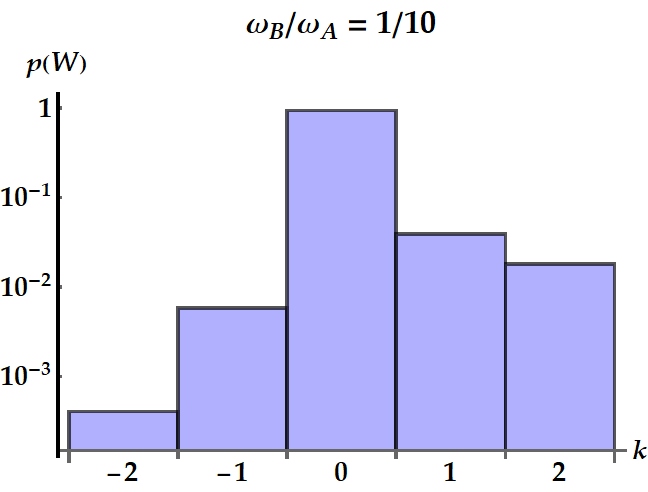} \\ \\
\includegraphics[scale=0.2]{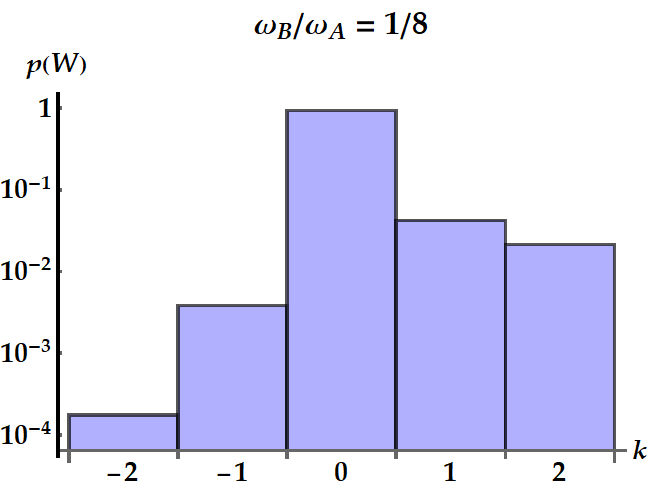} \hspace{0.2cm}
\includegraphics[scale=0.2]{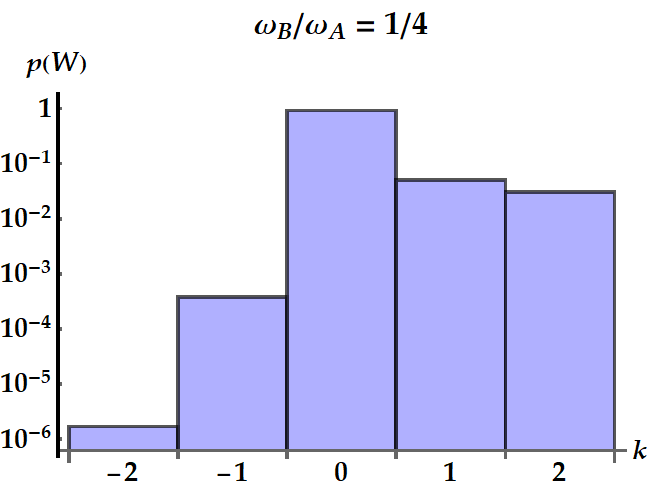} \hspace{0.2cm}
\includegraphics[scale=0.2]{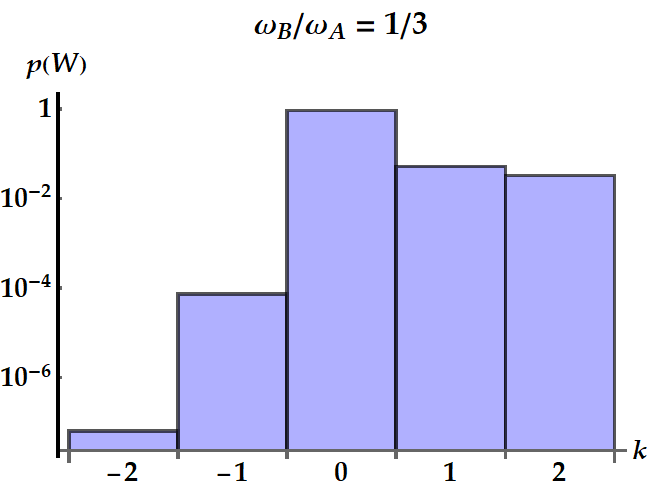}
\caption{Logarithmic plot of the distribution of work $p(W)$ given by the expansion up to the fourth order of $V_{12}$, with $\alpha = 0.5$,  $\beta_B/\beta_A=10^2$, $\beta_A\omega_A = 0.1$.}
\label{p2}
\end{figure}

In the following, as in the previous section, we fix $\theta = \sqrt{\alpha}\bar{\theta}$.
We show in Figs.~\ref{p} and~\ref{p2} the logarithmic plot of $p(W)$ for $V_{21}$ and $V_{12}$, respectively, for all values of $\omega_B/\omega_A$ in the sets outlined above, with $\beta_A\omega_A = 0.1$ and $\alpha = 0.5$. 
Differently from the moments of work and the efficiency, notice that the distribution of work does not depend on the bare frequencies, but just on $N_A$ and $N_B$.

\subsection{Average work} \label{3.2}
\begin{figure}
\centering
\includegraphics{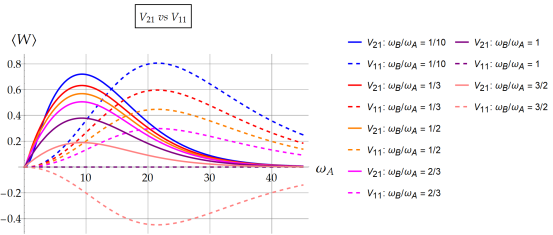}
\caption{Average work extracted by $V_{21}$ (solid curves) and by $V_{11}$ (dashed curves) with $\alpha = 0.5$, $\beta_B/\beta_A=10^2$ and different values of $\omega_B/\omega_A$ within the heat-engine regime of $V_{21}$.}
\label{V21vsV11}
\end{figure}

\begin{figure}
\centering
\includegraphics{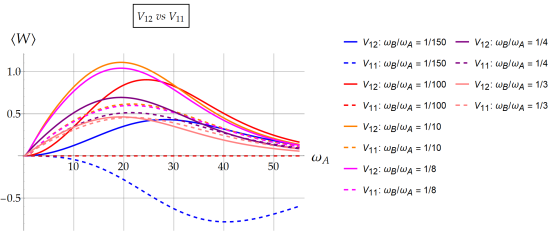}
\caption{Average work extracted by $V_{12}$ (solid curves) and by $V_{11}$ (dashed curves) with $\alpha = 0.5$, $\beta_B/\beta_A=10^2$ and different values of $\omega_B/\omega_A$ within the heat-engine regime of $V_{12}$.}
\label{V12vsV11}
\end{figure}

The mean work extracted by $V_{21}$ and $V_{12}$ is given by
\begin{eqnarray} \label{w21}
\langle W \rangle_{21} =& 2(2\omega_A-\omega_B)\alpha\bar{\theta}_{21}^2[N_A^2(N_B + 1) - (N_A + 1)^2N_B] \nonumber \\
&\left[1-\frac{2}{3}\alpha\bar{\theta}_{21}^2(6N_A - 4N_B + 1)\right] \\
\langle W \rangle_{12} =& 2(\omega_A-2\omega_B)\alpha\bar{\theta}_{12}^2[(N_B + 1)^2N_A - N_B^2(N_A + 1)] \nonumber \\
&\left[1-\frac{2}{3}\alpha\bar{\theta}_{12}^2(6N_B - 4N_A + 1)\right],
\end{eqnarray}
where we named the maximum couplings consistent with the distribution
of work as $\bar{\theta}_{21}$ and $\bar{\theta}_{12}$,
respectively. We plot the mean work with $\beta_A = 0.1$ as a function
of $\omega_A$ in Figs.~\ref{V21vsV11} and~\ref{V12vsV11}.
Numerically, we find that the maximum of $\langle W \rangle_{21}$ for fixed $\beta_A$ is obtained for $\omega_A \simeq 0.95/\beta_A$ and $x \simeq 0.08$: for this reason the blue curve for $x = 1/10$ in Fig.~\ref{V21vsV11} is the most performing in the plot. In the case of $\langle W \rangle_{12}$, the maximum for a fixed $\beta_A$ is found for $\omega_A \simeq 1.99/\beta_A$ and $x\simeq 0.03$. Note that the transformation $V_{12}$ can provide larger maxima of mean work with respect to $V_{21}$.
\\
It is interesting to compare $\langle W \rangle_{21}$ and $\langle W \rangle_{12}$ with the work extracted by the partial swap transformation $V_{11}$, namely \cite{max}
\begin{equation}
\langle W \rangle_{11} = (N_A - N_B) (\omega_A-\omega_B) \sin^2\theta.
\end{equation}
We provide a fair comparison of the performance of $V_{21}$ with
respect to $V_{12}$ by setting $\theta = \sqrt{\alpha}\bar{\theta}_{21}$ in Fig.~\ref{V21vsV11} and $\theta = \sqrt{\alpha}\bar{\theta}_{12}$ in Fig.~\ref{V12vsV11}. Notice that, for $x = 3/2$ in Fig.~\ref{V21vsV11} and $x = 1/150$ in Fig.~\ref{V12vsV11}, the partial swap does not extract work, while $V_{21}$ and $V_{12}$ do.

\begin{figure}
\centering
\includegraphics[scale=0.5]{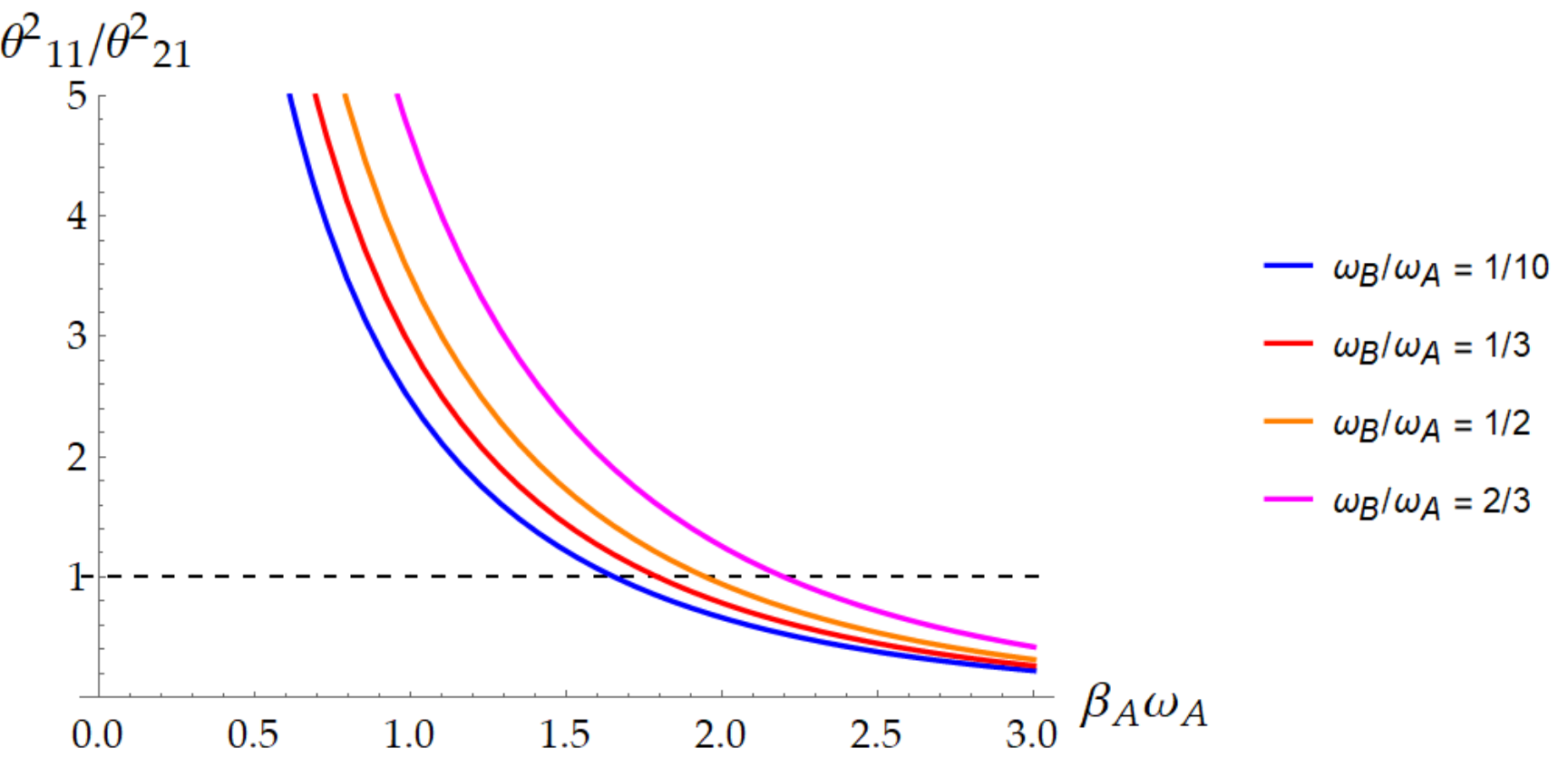} \hspace{0.5cm}
\includegraphics[scale=0.5]{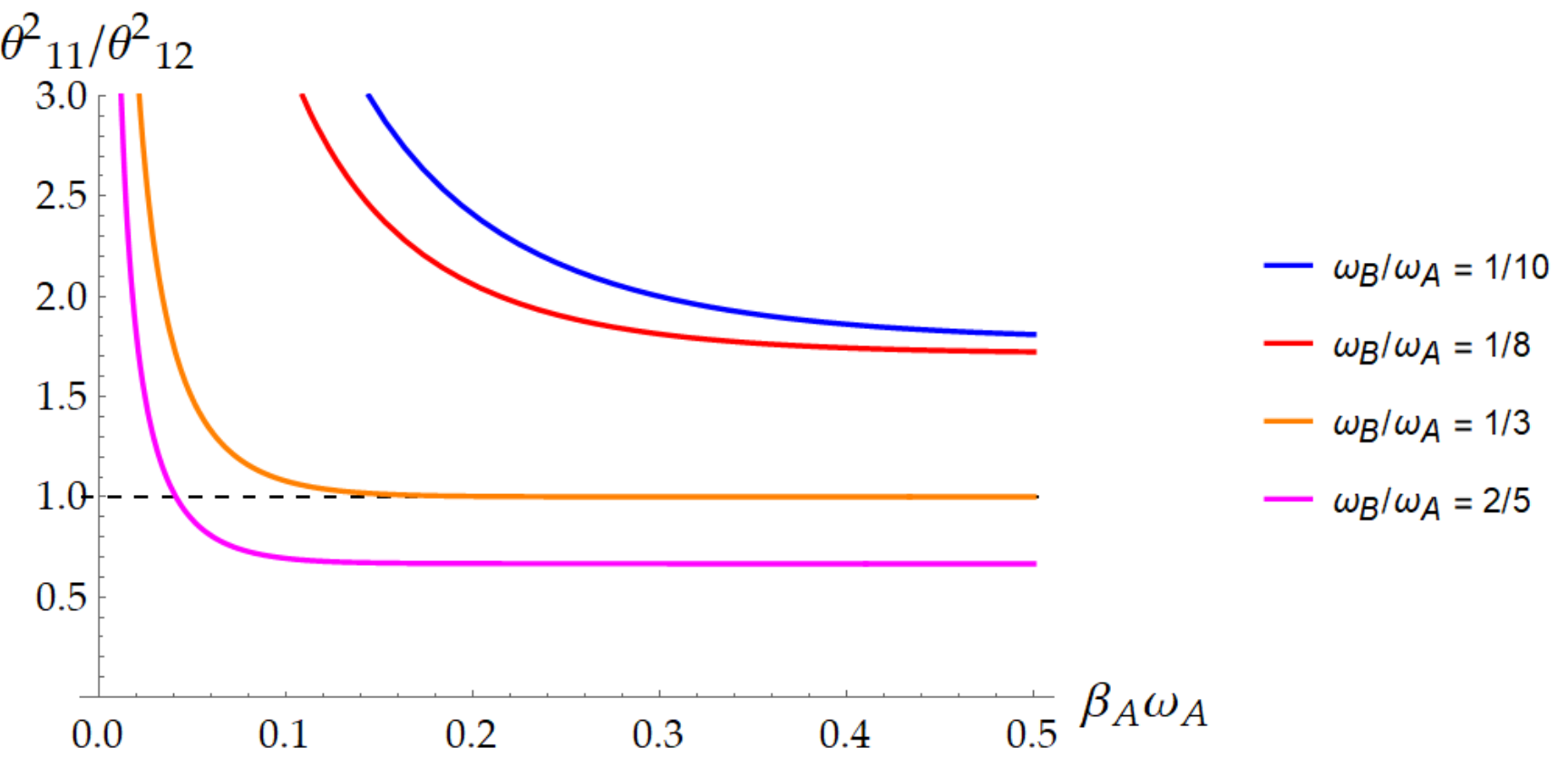}
\caption{Ratios $\theta_{11}^2/\theta_{21}^2$ in Eq.~(\ref{r21}) (left) and $\theta_{11}^2/\theta_{12}^2$ in Eq.~(\ref{r12}) (right) versus $\beta_A\omega_A$ with $\beta_B/\beta_A = 10^2$, for different values of the ratio $\omega_B/\omega_A$.}
\label{ratio}
\end{figure}

Now we let the couplings $\theta_{21}$, $\theta_{12}$ and $\theta_{11}$ free and look for the ratios $\theta^2_{11}/\theta^2_{21} \equiv r_{21}$ and $\theta^2_{11}/\theta^2_{12} \equiv r_{12}$ needed to obtain $\langle W_{11} \rangle = \langle W_{21} \rangle$ and $\langle W_{11} \rangle = \langle W_{12} \rangle$, respectively. Here we expand the interaction up to the second order. Ratios larger than 1 imply that the swap gate requires a larger coupling than the second-order interactions to achieve the same performance. Straightforward calculations yield
\begin{eqnarray}
r_{21} = \frac{2-x}{1-x}\frac{\sinh\left[\frac{\beta_A\omega_A}{2}(xy-2)\right]}{\sinh\left[\frac{\beta_A\omega_A}{2}(xy-1)\right]\sinh\left(\frac{\beta_A\omega_A}{2}\right)} \label{r21} \\
r_{12} = \frac{1-2x}{1-x}\frac{\sinh\left[\frac{\beta_A\omega_A}{2}(2xy-1)\right]}{\sinh\left[\frac{\beta_A\omega_A}{2}(xy-1)\right]\sinh\left(\frac{xy\beta_A\omega_A}{2}\right)}. \label{r12}
\end{eqnarray} 
We recall that, within our setting, $x = \omega_B/\omega_A$ and $y = \beta_B/\beta_A$ are fixed, so that $\beta_B\omega_B = xy \beta_A\omega_A$. If $\beta_A\omega_A < (xy)^{-1}$, implying $\beta_B\omega_B < 1$, then $r_{21} \propto (\beta_A\omega_A)^{-1}$, while for $\beta_A\omega_A \gg (xy)^{-1}$  the ratio of the couplings drops as $r_{21}\propto (e^{\beta_A\omega_A}-1)^{-1} = N_A$.  We see this last behavior in the left panel of Fig.~\ref{ratio}, where we plot $r_{21}$ as a function of $\beta_A\omega_A$ for four values of $\omega_B/\omega_A$. Note that the two unitaries extract the same work with the same coupling, identified by the black dashed line, for $\beta_A\omega_A \simeq 2$. The coupling $\theta_{11}$ of the swap needs to be stronger for $\beta_A\omega_A < 2$ and weaker  for $\beta_A\omega_A > 2$. This result is consistent with Fig.~\ref{V21vsV11} corresponding to $r_{21} = 1$ and $\beta_A = 0.1$, where one can see that $\langle W \rangle_{21} > \langle W \rangle_{11}$ for $\omega_A \lesssim 2/\beta_A = 20$. There are discrepancies between the intersection points of $\langle W \rangle_{21}$ and $\langle W \rangle_{11}$ in Fig.~\ref{V21vsV11} and the corresponding points with $r_{21} = 1$ in Fig.~\ref{ratio}, due to the different order of the perturbative expansion explored in the two plots.
\\
The case of $r_{12}$, plotted in the right panel of Fig.~\ref{ratio}, is even more interesting. In the regime $\beta_A\omega_A \gg (xy)^{-1}$, now we have $r_{12} \rightarrow 2(1-2x)/(1-x)$, implying that $V_{12}$ outperforms $V_{11}$ in terms of mean extracted work $\forall \, x < 1/3$, consistently with Fig.~\ref{V12vsV11}. Here, differently from the previous case, one can choose $x$ and $y$ such that $V_{12}$ outperforms the swap gate for every value of $\beta_A\omega_A$.

\subsection{Signal-to-noise ratio} \label{3.3}
\begin{figure}
\includegraphics[scale=0.22]{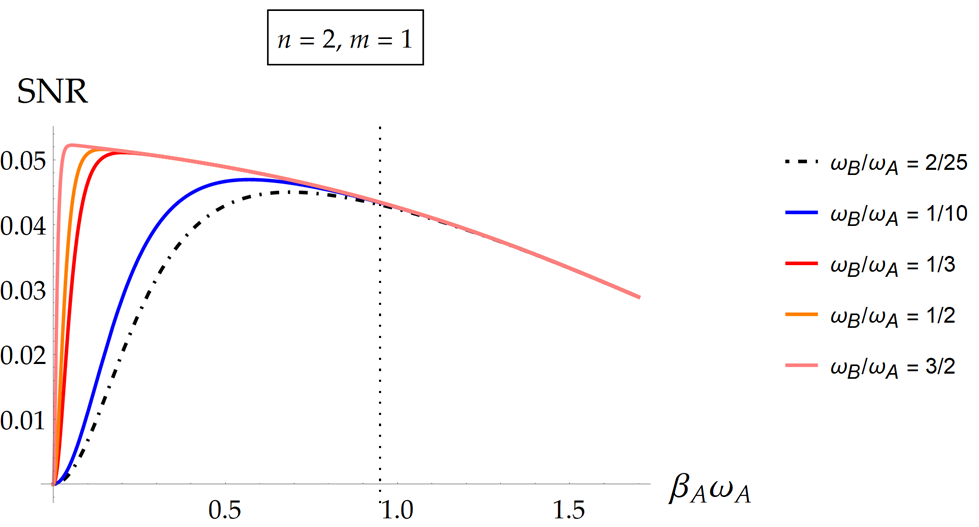} \hspace{0.25cm}
\includegraphics[scale=0.22]{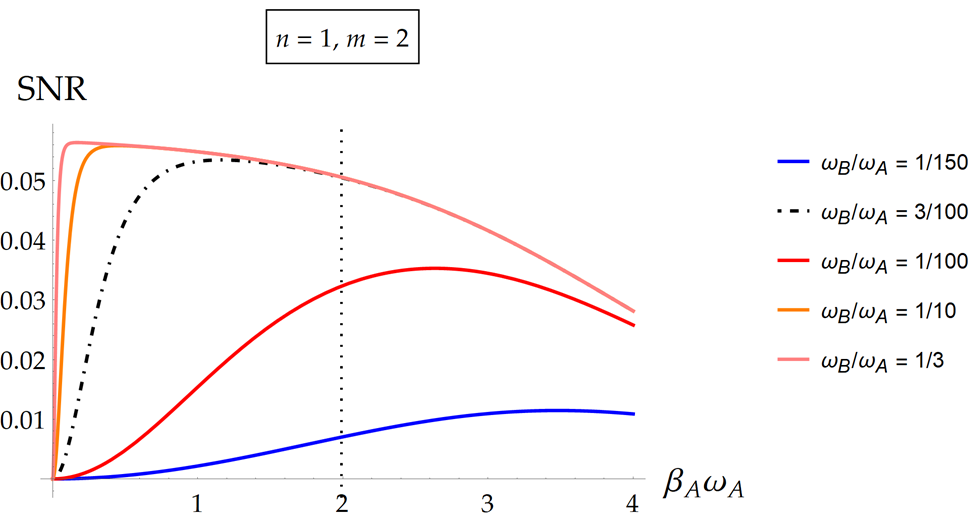} 
\caption{SNRs for $V_{21}$ in Eq.~(\ref{snr21}) (left) and for $V_{12}$ in Eq.~(\ref{snr12}) (right) with $\beta_B/\beta_A=10^2$, $\alpha = 0.5$. The dot-dashed black curves display the SNR for the value of $\omega_B/\omega_A$ maximizing the average work, whose maximum over $\beta_A\omega_A$ is identified by the vertical black dotted lines.}
\label{snr}
\end{figure}

Finally, we investigate the SNRs of the work extracted by $V_{21}$ and $V_{12}$. First, we relate the SNRs to the mean entropy production $\langle \Sigma \rangle$ in Eq.~(\ref{mep}). Up to the fourth order, the second moment of the extracted work reads
\begin{equation} \label{w2_21}
\langle W^2 \rangle_{21} \sim 2\alpha\bar{\theta}_{21}^2(2\omega_A - \omega_B)^2[\mathcal{A} -  2\alpha\bar{\theta}^2_{21}(\mathcal{A}\mathcal{B} - 4\mathcal{C})],
\end{equation}
for the unitary $V_{21}$, with $\mathcal{A}$, $\mathcal{B}$ and $\mathcal{C}$ defined in Eqs.~(\ref{A}),~(\ref{B}) and~(\ref{C}), and
\begin{equation}
\langle W^2 \rangle_{12} \sim 2\alpha\bar{\theta}_{12}^2(2\omega_B - \omega_A)^2[\widetilde{\mathcal{A}} -  2\alpha\bar{\theta}^2_{12}(\widetilde{\mathcal{A}}\widetilde{\mathcal{B}} - 4\widetilde{\mathcal{C}})]
\end{equation}
for $V_{12}$, where $\widetilde{\mathcal{A}}$, $\widetilde{\mathcal{B}}$ and $\widetilde{\mathcal{C}}$ are obtained from $\mathcal{A}$, $\mathcal{B}$ and $\mathcal{C}$ by exchanging $N_A$ with $N_B$.
\\
Hence, from Eqs.~(\ref{mep}),~(\ref{w21}) and~(\ref{w2_21}), we retrieve the SNR  for $V_{21}$ in terms of the mean entropy production as follows
\begin{eqnarray}
{\rm SNR}_{21} &\equiv \frac{\langle W \rangle^2_{21}}{\langle W^2 \rangle_{21} - \langle W \rangle^2_{21}} \nonumber \\
& = \frac{\langle \Sigma \rangle_{21}(2\omega_A - \omega_B)}{\beta_B\omega_B-2\beta_A\omega_A}\frac{\langle W \rangle_{21}}{\langle W^2 \rangle_{21} - \langle W \rangle^2_{21}} \nonumber \\
& =\langle \Sigma \rangle_{21} \frac{\tanh\left(\frac{\beta_B\omega_B-2\beta_A\omega_A}{2}\right)}{\beta_B\omega_B-2\beta_A\omega_A}(1-\alpha\bar{\theta}^2_{21}\Delta)^{-1} \label{snr21}
\end{eqnarray}
where
\begin{equation}
\Delta \equiv 1 - \frac{\mathcal{C}}{\mathcal{B}\mathcal{A}} - \frac{\mathcal{D}^2}{\mathcal{B}\mathcal{A}},
\end{equation}
with
\begin{equation}
\mathcal{D} \equiv N_A^2(N_B+1) - (N_A+1)^2N_B.
\end{equation}
Similarly, for $V_{12}$ one finds
\begin{equation}
{\rm SNR}_{12} = \langle \Sigma \rangle_{12} \frac{\tanh\left(\frac{2\beta_B\omega_B-\beta_A\omega_A}{2}\right)}{2\beta_B\omega_B-\beta_A\omega_A}(1-\alpha\bar{\theta}^2_{12}\widetilde{\Delta})^{-1} \label{snr12}
\end{equation}
where $\widetilde{\Delta}$ is obtained from $\Delta$ by exchanging
$N_A$ and $N_B$.  \\ At the second order, we recover the reciprocal of
the RFs outlined in Eq.~(\ref{rf2ord}), constrained by the standard
TUR in Eq.~(\ref{tur}). This term of the expansion in the SNR is
proportional to $\theta^2$ through $\langle \Sigma \rangle$ and
determines the dependence of the SNR on $\beta_A\omega_A$ and
$\omega_B/\omega_A$, shown in Fig.~\ref{snr} for $V_{21}$ (left panel)
and $V_{12}$ (right panel). For $\beta_B\omega_B \gg 2\beta_A\omega_A$
[$\beta_B\omega_B \gg \beta_A\omega_A/2$] the hyperbolic tangent in
Eq.~(\ref{snr21}) [Eq.~(\ref{snr12})] approaches unity and the SNR
does not depend on $\omega_B/\omega_A$. Conversely, in the opposite
regime, the SNR grows monotonically with
$\omega_B/\omega_A$, since $\langle \Sigma \rangle_{21} \propto
\beta_B\omega_B - 2\beta_A\omega_A$ (and similarly for $\langle \Sigma
\rangle_{12}$) and the tangent is approximated by its argument. The
maximum of the SNR is achieved for $\omega_B = 2\omega_A$, in the case
of $V_{21}$, and $\omega_A = 2\omega_B$, in the case of $V_{12}$,
where no work is extracted. However, for both unitaries, it is
possible to extract maximum mean work with non-negligible SNR, as
highlighted by the dot-dashed black curves, identifying the SNRs for
the value of $\omega_B/\omega_A$ optimizing the average work, and by
the vertical dotted lines, that select the corresponding optimal
$\beta_A\omega_A$.

\begin{figure}
\includegraphics[scale=0.53]{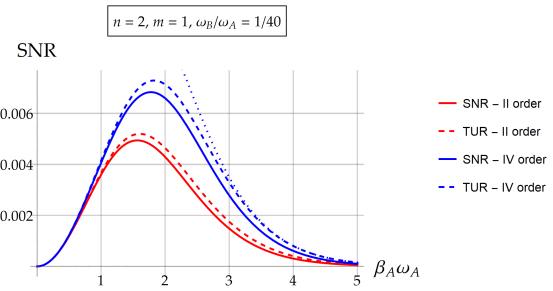} \hspace{0.5cm}
\includegraphics[scale=0.53]{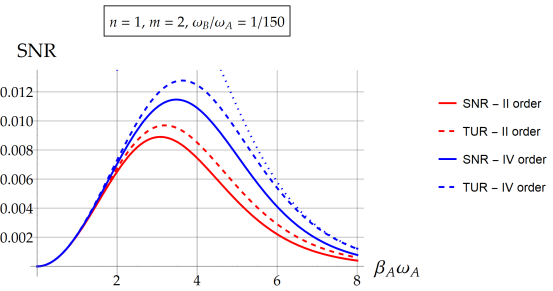} \hspace{0.5cm}
\caption{SNRs for $V_{21}$ (left) and for $V_{12}$ (right), up to the second (red curves) and fourth (blue curves) order in $\theta$, with $\alpha = 0.5$ and $\beta_B/\beta_A=10^2$. The dashed blue curves display the upper bound for the TURs at the fourth order in $\theta$ in Eqs.~(\ref{ub1}), in the left panel, and~(\ref{ub2}), in the right panel. The dashed red curves are the standard TURs in Eq.~(\ref{tur}). The dotted blue curve is the asymptotic value of the upper bounds in the TUR at the fourth order in $\theta$, namely Eqs.~(\ref{ub11}), in the left panel, and~(\ref{ub22}), in the right panel.}
\label{snrx}
\end{figure}

The fourth order is interesting for what concerns the TURs. Indeed, the SNRs for $V_{21}$ and $V_{12}$ are upper bounded by
\begin{eqnarray}
{\rm SNR}_{21} \leq \frac{\langle \Sigma \rangle_{21}}{2} (1-\alpha\bar{\theta}^2_{21}\Delta)^{-1} \label{ub1} \\
{\rm SNR}_{12} \leq \frac{\langle\Sigma \rangle_{12}}{2} (1-\alpha\bar{\theta}^2_{12}\widetilde{\Delta})^{-1} \label{ub2}
\end{eqnarray}
implying that a violation of the pertaining standard TURs in
Eq.~(\ref{tur}) can be observed for positive $\Delta$ or
$\widetilde{\Delta}$. This is the case for both $V_{21}$ and $V_{12}$,
since, for every value of $\omega_B/\omega_A$ here considered, numeric
evaluations show that $\Delta > 0$ if $\beta_A\omega_A \gtrsim 1.24$
and $\widetilde{\Delta} > 0$ if $\beta_A\omega_A \gtrsim
3.26$. Moreover, for $\omega_B/\omega_A$ approaching the inferior
boundary of the operational regime, namely for $\omega_B/\omega_A
\rightarrow (n\beta_A)/(m\beta_B)$, both $\Delta$ and
$\widetilde{\Delta}$ are positive for any value of
$\beta_A\omega_A$. We remark that the fourth order rescales the upper
bound given by the standard TUR, ultimately yielding new TURs. We show this behavior in Fig.~\ref{snrx}, where the solid lines display the SNRs up to the second (in red) and fourth (in blue) order; the blue dashed lines identify the upper bounds in Eqs.~(\ref{ub1}) for $V_{21}$ (left panel), and~(\ref{ub2}) for $V_{12}$ (right panel), while the dashed red curves are the upper bounds given by the standard TURs in Eq.~(\ref{tur}). It is worth noting that, for $\beta_A\omega_A \gg 1$, both $\bar{\theta}^2_{21}\Delta$ and $\bar{\theta}^2_{12}\widetilde{\Delta}$ approach unity, thus simplifying the TURs as
\begin{eqnarray}
{\rm SNR}_{21} \leq \frac{\langle \Sigma \rangle_{21}}{2} (1-\alpha)^{-1} \label{ub11} \\
{\rm SNR}_{12} \leq \frac{\langle\Sigma \rangle_{12}}{2} (1-\alpha)^{-1} \label{ub22}.
\end{eqnarray}
These asymptotic bounds are also reported, as blue dotted curves, in the two panels of Fig.~\ref{snrx}. 

\begin{figure}
\includegraphics{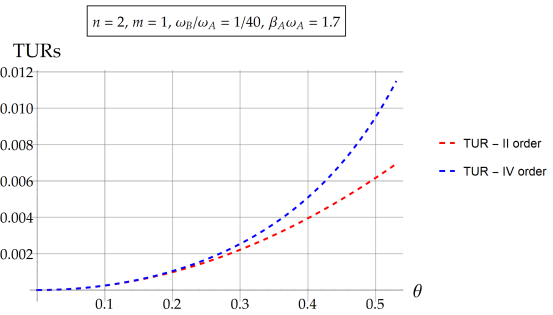}
\caption{Upper bounds in the TUR in Eq.~(\ref{ub1}) (blue curve) and in the standard TUR in Eq.~(\ref{tur}) for $\beta_A\omega_A = 1.7$, $\omega_B/\omega_A = 1/40$ and $\beta_B/\beta_A=10^2$.}
\label{thetanoass}
\end{figure}

Surprisingly, the fourth order improves considerably the SNR in this
scenario. On the one hand, the conditions imposed by the distribution
of work ($0\leq p(W) \leq 1$) do not constraint the coupling $\theta$
to infinitesimal values, without making it incompatible with the
expansion of the unitaries ($\theta < 1$). On the other hand, one may
argue that this effect could be a consequence of our assumption
$\theta = \sqrt{\alpha}\bar{\theta}$, since it introduces an
artificial dependence of $\theta$ on $N_A$ and $N_B$. We show that
this is not the case: the effect in Fig.~\ref{snrx} is still present
as a function of $\theta$. Let us consider the case of $V_{21}$ in the
left panel of Fig.~\ref{snrx} and take $\beta_A\omega_A \simeq 1.7$
such that both the TURs (dashed lines) at the second and fourth order
give a large value. We plot in Fig.~\ref{thetanoass} the upper bounds
given by the standard TUR (red curve) and the TUR at the fourth order
in Eq.~(\ref{ub1}) (blue curve), where $\theta$ is now left free, thus
confirming the improvement found at the fourth order evaluation.
\\
We emphasize that the relaxation of the TURs at fourth order, as expressed in Eqs.~(\ref{ub11}) and~(\ref{ub22}), arises from the same mechanism as the second-order case - namely, the finite-order expansion of the interaction $V_{\theta}$. These bounds are therefore realistic under the assumption that contributions of order $\theta^6$ and higher are negligible. In our analysis, this assumption is enforced by constraining $\theta$ to remain below the threshold $\bar{\theta}$ defined in Eq.~(\ref{thetaup2}).  It is important to note that no definitive conclusions can be drawn from our analysis regarding the form of TURs that might bound the SNRs when the full, non-perturbative expansion of the interaction is considered. Indeed, even in the case of the swap gate $V_{11}$ for which the exact dynamics is known, a similar perturbative relaxation of the TUR can be observed. Nevertheless, we know that the stricter TUR given in Eq.~(\ref{maxtur}) holds exactly for arbitrarily large coupling in this case, underscoring the limitations of low-order expansions in capturing the full thermodynamic constraints.

\section{Conclusions} \label{4}
We have provided an extensive analysis of two-stroke heat engines
where the working fluid is represented by two bosonic modes coupled by
a polynomial nonlinear interaction. In particular, we have addressed
the general FC unitary in Eq.~(\ref{fc}) by a second order expansion
in the coupling $\theta$.
 
By adopting the two-point measurement scheme we have obtained the
distribution for the stochastic work and heat, and hence we have studied
the relative fluctuations of the extracted work up to the second order
in the coupling $\theta$ for different choices of $n$ and $m$.

We have shown that for small values of the ratio $\beta_B/\beta_A$ one
does not need to exploit extremely high orders, since the maximum of
the average extracted work is achieved for small $n$ and $m \leq
n$. On the contrary, for large $\beta_B/\beta_A$ the optimal choice is
obtained by using the highest $n$ achievable and $m < n$, which also
allows minimal RFs.  \\ Then, we have inspected the specific
interactions $V_{21}$ and $V_{12}$ up to the fourth order in $\theta$,
by assuring the consistency of the Taylor expansion with suitable
constraints on the coupling.  In terms of average work, we have shown
that both unitary operations outperform the swap gate, by also
enlarging the regime of operation of the heat engine. In particular,
for $\beta_B/\beta_A = 10^2$ and $\omega_B/\omega_A < 1/3$, the
unitary $V_{12}$ extracts a larger amount of work for all values of
$\beta_A\omega_A$. Furthermore, for both $V_{21}$ and $V_{12}$, the
extraction of work is compatible with non-negligible SNRs, even in the
case where the average work is maximized. Finally, we have identified
the TURs that bound the SNRs for these interactions at the fourth
order in $\theta$, by showing that they allow for larger SNRs than the
standard TURs retrieved at the second order. In fact, by this final
result, we have provided the conditions determining a violation of the
standard TURs by the interactions $V_{21}$ and $V_{12}$ and have
presented specific cases where this effect is clearly visible.

\ack{Giovanni Chesi and Chiara Macchiavello acknowledge funding from the PNRR MUR Project PE0000023-NQSTI. Massimiliano Federico Sacchi acknowledges support from the PRIN MUR Project 2022SW3RPY.}

\end{document}